\documentclass[11pt]{article}

\usepackage[T1]{fontenc}
\usepackage[margin=1in]{geometry}

\usepackage{mathtools}
\usepackage{amsmath}
\usepackage{amsfonts}
\usepackage{amssymb}
\usepackage{amsthm}
\usepackage{bm}
\usepackage{graphicx}
\usepackage{algorithm}
\usepackage{algpseudocode}
\usepackage{tikz}
\usepackage{xspace}
\usepackage{enumitem}
\usepackage{thmtools,thm-restate}
\usepackage[round]{natbib}
\usepackage{endnotes}
\usepackage{hyperref}
\usepackage{cleveref}
\crefname{assumption}{Assumption}{Assumptions}
\Crefname{assumption}{Assumption}{Assumptions}

\newtheorem{assumption}{Assumption}
\theoremstyle{definition}
\newtheorem{definition}{Definition}
\newtheorem{example}{Example}

\theoremstyle{remark}
\newtheorem{remark}{Remark}

\newcommand{\sppe}{\textsf{SPPE}\xspace}
\newcommand{\fppe}{\textsf{FPPE}\xspace}
\newcommand{\ppad}{\textsf{PPAD}\xspace}

\newcommand{\R}{\mathbb{R}}
\newcommand{\E}{\mathbb{E}}
\newcommand{\B}{\mathcal{B}}
\newcommand{\BR}{\mathrm{BR}}
\newcommand{\one}{\mathbf{1}}
\newcommand{\ind}{\mathbf{1}}
\newcommand{\norm}[1]{\lVert #1\rVert}
\newcommand{\abs}[1]{\lvert #1\rvert}
\newcommand{\dd}{\,\mathrm{d}}
\DeclareMathOperator{\diag}{diag}

\title{Convergence of Pacing Dynamics in Second-Price Auction Markets}

\author{
\begin{tabular}{ccc}
Salam Afiouni & Christian Kroer  \\[2pt]
Columbia University & Columbia University\\
\texttt{sa4316@columbia.edu} & \texttt{christian.kroer@columbia.edu}
\end{tabular}
}

\date{}

\begin{document}

\maketitle

\begin{abstract}
Computing an approximate second-price pacing equilibrium (\sppe) is \ppad-complete, and \ppad-hardness persists even for constant-factor approximations. Yet budget pacing is widely used in online advertising markets. We identify market conditions under which \sppe recovers tractability.
We study fluid second-price auction markets and show that a new uniform \emph{own-spend dominance} condition, which requires a bidder's spending to be more sensitive to her own pacing multiplier than to rivals' combined influence, ensures a unique equilibrium and contraction of best-response dynamics. When this condition holds throughout the pacing domain, the widely used dual descent approach is also contractive and stochastic pacing converges almost surely.

Motivated by heterogeneity in click-through rates across advertising impressions, we then consider a logit valuation model and derive sufficient conditions for own-spend dominance in terms of win shares and spending shares. These include a large-market condition under which sufficiently many comparable rivals on each item type guarantee uniqueness and convergence.

Finally, we extend the analysis to finite-item markets, establishing high-probability guarantees for inverse-polynomial approximate equilibria in randomly sampled markets and polynomial-time computability for a class of fluid-approximable instances.
\end{abstract}

\section{Introduction}\label{sec:intro}

Online auctions play a central role in the allocation of advertising opportunities. Whenever a user visits a webpage or submits a search query, advertisers may compete in an auction for the opportunity to display an advertisement. An advertiser typically participates in thousands or millions of such auctions over the course of a campaign while facing a hard budget constraint on her total expenditure. In practice, advertisers do not submit bids in individual auctions directly. Instead, they specify campaign-level parameters, such as their value per click, budget, and targeting criteria, and delegate auction-level bidding decisions to the platform. In effect, the platform operates a proxy bidder on behalf of each advertiser, determining which auctions she participates in, how aggressively she bids, and how her budget is allocated across advertising opportunities. Budget management is therefore a central component of automated bidding systems.

A widely used approach to budget management is \emph{multiplicative pacing} \citep{meta_pacing, linkedinhelp_talentinsights_a422101,adobe2026pacing}, whereby each bidder's valuations are scaled by a common factor, a \emph{pacing multiplier}, which takes values in $[0,1]$. The pacing multiplier is strictly smaller than one only if the bidder would exceed her budget by bidding her true value in each auction. By adjusting a single parameter for each participant, the platform can regulate expenditure across a large number of heterogeneous auctions, preventing bidders from exhausting their budgets prematurely and thereby missing valuable advertising opportunities later in the campaign. 

Beyond its practical appeal, multiplicative pacing admits a particularly natural economic and strategic interpretation in second-price auctions. Holding the bids of competing bidders fixed, a bidder's own bid determines whether she wins an item but not the price she pays upon winning it. Multiplicative pacing therefore induces a threshold on the value-to-price ratio of the items she acquires, allowing her to prioritize opportunities that yield the highest return per unit of expenditure. Under appropriate assumptions on allocation and tie-breaking, a budget-constrained bidder's best response can thus be implemented through a single pacing multiplier \citep{babaioff2020non}. From a dual perspective, this multiplier is determined by the shadow price of the bidder's budget constraint~\citep{balseiro2015repeated,balseiro2019learning}. The optimality of pacing extends beyond the static setting. For a budget-constrained bidder participating in repeated second-price auctions, with valuations drawn i.i.d. from an unknown distribution, \citet{balseiro2019learning} establish asymptotic optimality guarantees for adaptive pacing under both stochastic and adversarial competition, achieving optimal-order regret in the former setting and an optimal asymptotic competitive ratio in the latter.

In markets where all bidders employ multiplicative pacing, each bidder's expenditure depends not only on her own pacing multiplier but also on those of her competitors. This interdependence motivates the notion of a \emph{pacing equilibrium}, introduced and studied in first- and second-price auction markets by \citet{conitzer2022pacing,conitzer2022multiplicative}. A pacing equilibrium consists of a profile of pacing multipliers and an associated allocation such that items are allocated according to the auction rules, all bidders respect their budget constraints, and no bidder is unnecessarily paced: a bidder who does not exhaust her budget must have a pacing multiplier equal to one. 

In second-price settings, pacing equilibria admit a natural strategic interpretation as Nash equilibria of the underlying budget-constrained bidding game under appropriate assumptions on budget enforcement and unilateral deviations~\citep{balseiro2015repeated,balseiro2019learning,conitzer2022multiplicative,babaioff2020non}. However, their structural and computational properties are less favorable: second-price pacing equilibria (\sppe) need not be unique, and can violate natural monotonicity and optimality properties \citep{conitzer2022multiplicative}. Moreover, \citet{chen2024complexity} show that computing an approximate \sppe is \ppad-complete, and \citet{chen2025constant} further show that \ppad-hardness persists even for constant-factor approximations. Thus, despite the strong theoretical justification for multiplicative pacing at the individual level, its simultaneous adoption by multiple bidders can give rise to a computationally intractable equilibrium problem. 
This stands in sharp contrast to first-price auction markets, where equilibrium pacing multipliers are unique and can be computed efficiently through an Eisenberg-Gale convex program \citep{conitzer2022pacing}. These favorable properties extend to a broader class of first-price-like mechanisms under suitable conditions~\citep{afiouni2026pacing}.

In practice, advertising platforms employ feedback-based pacing algorithms that dynamically adjust bidding parameters in response to observed spending and budget targets \citep{xie2021twitterpacing,meta2026auction,linkedin2026pacing}.
Given the computational barriers discussed above, a natural question is under what conditions such dynamics converge to a pacing equilibrium.
In an early contribution, \citet{borgs2007dynamics} propose a t\^{a}tonnement-style process in which bidders adjust their bids in response to deviations of their spending from their budget targets. They establish convergence results for first-price auctions and provide numerical evidence of convergence in second-price auctions. For the latter, \citet{balseiro2019learning} establish convergence when all bidders simultaneously employ adaptive pacing strategies, under a strong monotonicity assumption on the expected spending map. In general, however, even best-response dynamics can cycle \citep{conitzer2022multiplicative}, and the aforementioned computational hardness results preclude general polynomial-time convergence guarantees to approximate pacing equilibria, unless $\ppad=\textsf{P}$. More recently, \citet{gaitonde2022budget} establish individual regret and liquid welfare guarantees for pacing dynamics without requiring convergence to equilibrium.

The convergence properties of pacing dynamics are closely related to how bidders' spending responds to changes in pacing multipliers. In second-price auctions, changes in competing bids can affect a bidder's spending even when they leave her allocation unchanged. For instance, a bidder who consistently outbids her competitors by a large margin may continue to win the same items as her competitors increase their pacing multipliers, while paying higher prices for these items. By contrast, changes in her own multiplier have no effect on her spending as long as her allocation remains unchanged. Indeed, the ability of a bidder to influence another bidder's spending without effectively competing for the allocation plays an important role in the hardness construction of \citet{chen2024complexity}. This motivates the central question of this paper:

\begin{quote}
\emph{Under what conditions on competition in second-price auction markets is the pacing equilibrium unique, and do simple pacing dynamics converge to it?}
\end{quote}

We address this question by studying the relationship between the structure of market competition and the sensitivity of bidders' spending to pacing adjustments. When bidders compete closely for the same items, the increase in payments induced by higher competing bids may be offset by a reduction in spending due to lost allocations. The interplay between these two effects is central to our analysis.

Motivated by the large number of heterogeneous auction opportunities in online advertising markets, we begin with a fluid model of second-price auctions in which spending is aggregated over a continuum of items. This formulation allows us to characterize the marginal effects of pacing adjustments on spending and establish sufficient conditions for uniqueness and convergence. We then study these conditions in a logit valuation model, relating them to economically interpretable measures of market competition. Finally, we extend our analysis to finite-item markets, where spending is generally discontinuous in the pacing multipliers.

\subsection{Contributions}

We establish sufficient conditions for the uniqueness of second-price pacing equilibria and the convergence of pacing dynamics, characterize these conditions in terms of market competition, and derive implications for equilibrium computation in finite auction markets.
\begin{itemize}
    \item \textbf{Own-spend dominance and convergence of pacing dynamics.}
    We first study a fluid model of second-price auctions, in which bidders' spending is aggregated over a continuum of items. We introduce an \emph{own-spend dominance} condition, which requires each bidder's spending to be more sensitive to her own log pacing multiplier than to the combined effects of her competitors' log multipliers. When this condition holds uniformly at profiles where bidders exhaust their budgets, we establish uniqueness of the \sppe and contraction of the best-response map, implying convergence of both synchronous and asynchronous best-response dynamics. When own-spend dominance holds throughout the pacing domain, we further establish contraction of projected dual descent and almost-sure convergence of stochastic pacing dynamics based on realized auction payments.
    
    \item \textbf{The role of marginal competition.} 
    We characterize own-spend dominance in terms of two opposing effects of competing bids on spending: a \emph{price effect}, which increases payments on retained items, and a \emph{displacement effect}, which reduces spending through lost allocations. We show that own-spend dominance holds precisely when the aggregate price effect that is not offset by displacement accounts for less than half of the bidder's spending. This characterization formalizes the notion of \emph{marginal competition} and provides an economic interpretation of our convergence conditions. By contrast, under first-price payments, own-spend dominance holds automatically whenever spending is positive.

    \item \textbf{Convergence in logit markets.}
    Motivated by impression-level heterogeneity in online advertising, we consider a logit valuation model in which price and displacement effects admit closed-form expressions. We derive sufficient conditions for own-spend dominance in terms of bidders' win shares and spending shares. In particular, we show that sufficiently many competitors with comparable bid levels on each item type guarantee uniqueness of the \sppe and convergence of pacing dynamics. The sufficient condition thus concerns the number of comparable competitors on each item type, rather than the total number of bidders in the market.

    \item \textbf{From fluid markets to finite instances.}
    We extend our analysis to finite-item markets by showing that best-response dynamics approximately inherit the contraction properties of a sufficiently close fluid market satisfying suitable dominance and regularity conditions. For randomly sampled markets with bounded valuations, we obtain high-probability guarantees for computing inverse-polynomial approximate \sppe. More generally, we identify a class of fluid-approximable instances for which an approximate \sppe can be computed in polynomial time, without knowledge of the underlying fluid market. 
\end{itemize}

\subsection{Pacing Dynamics in Practice}

We now describe in more detail how pacing operates on advertising platforms and how our model relates to this setting. Repeated second-price auctions with budget pacing are used by major platforms \citep{linkedin2026auction,linkedin2026pacing,x2026bidding}. Consider an advertiser who derives value from clicks on her advertisement. When an impression becomes available, the proxy bidder computes her expected value for the impression as the product of her value per click and the predicted click-through rate \citep{gao2022bidding,richardson2007predicting,mcmahan2013ad,he2014practical}. Since predicted click-through rates depend on the user, the context, and the advertisement, valuations vary across impressions, and the same advertiser may face strong competition for some impressions and little competition for others. Under multiplicative pacing, the proxy bidder submits this value scaled by the advertiser's current pacing multiplier.

In practice, budget management often involves two components: an expenditure plan that distributes the advertiser's budget over the duration of the campaign and a pacing controller that tracks this plan \citep{balseiro2023robust}. Under multiplicative pacing, the controller decreases the pacing multiplier when expenditure exceeds its target and increases it when expenditure falls below its target, subject to the upper bound of one. Such feedback-based adjustment procedures are discussed by \citet{gao2022bidding} in the context of LinkedIn's automated bidding system and by \citet{xie2021twitterpacing} in the context of Twitter's pacing service. Although pacing adjustments can be performed using only campaign-specific spending information, each campaign's spending depends on its competitors' bids. These controllers therefore interact through the auctions and jointly give rise to a dynamical system. Our model abstracts from the problem of determining expenditure plans and studies the behavior of interacting pacing controllers under fixed spending targets and stationary market conditions. A pacing equilibrium corresponds to a steady state of the adjustment dynamics considered in our model.

The dynamics we study capture two forms of such feedback-based control: projected dual descent adjusts pacing multipliers based on expected spending, whereas per-auction stochastic pacing adjusts them based on each bidder's own realized payments. Our results identify conditions on the structure of market competition under which both dynamics converge to a unique pacing equilibrium. These results provide a possible explanation for why pacing controllers perform well in practice despite the worst-case hardness of second-price pacing. At the same time, our analysis suggests that pacing dynamics may become unstable in market segments where an advertiser consistently wins impressions by a large margin and pays prices determined by much weaker competitors.
\section{Model}\label{sec:model}

\subsection{Second-Price Auctions and Pacing Equilibria}

Consider $n\ge2$ bidders and a measurable item space $(\Theta,\mathcal A)$ endowed with a finite supply measure~$\mu$. Each bidder $i$ has a nonnegative valuation function $v_i:\Theta\to\R_{\ge0}$ with finite, positive total value $V_i=\int_\Theta v_i(\theta)\,\dd\mu(\theta) \in(0,\infty),$ and a budget $B_i>0$. The finite-item model with $m$ items is recovered by taking $\Theta$ to be a set of $m$ points and $\mu$ to be a counting measure.

Each bidder scales her values by a common \emph{pacing multiplier} $\alpha_i\in(0,1]$. We work with log multipliers $\beta_i=\log\alpha_i\le0$, so bidder $i$'s bid on item $\theta$ is $b_i(\theta;\beta)=e^{\beta_i}v_i(\theta).$ The highest bid and the second-price payment on item $\theta$ are, respectively,
\[
h(\theta;\beta)=\max_i b_i(\theta;\beta),
\qquad
p(\theta;\beta)=
\text{second-highest of } \{b_i(\theta;\beta)\}_i.
\] An allocation is a measurable map $x:\Theta\to[0,1]^n$, where $x_i(\theta)$ is the fraction of item $\theta$ allocated to bidder $i$. Bidder $i$'s total spending under allocation $x$ is $S_i(\beta;x) =\int_\Theta x_i(\theta)\, p(\theta;\beta)\,\dd\mu(\theta).$

\begin{definition}[SPPE, \citealp{conitzer2022multiplicative,chen2024complexity}]\label{def:sppe}
A pair $(\beta,x)$ is a \emph{second-price pacing equilibrium} (\sppe) if the following conditions hold for every bidder, with the itemwise conditions understood $\mu$-almost everywhere:
\begin{enumerate}[leftmargin=2em,label=(\alph*)]
\item \emph{Highest bids win:} $x_i(\theta)>0$ implies $b_i(\theta;\beta)=h(\theta;\beta)$.
\item \emph{Items with positive bids are sold:} $h(\theta;\beta)>0$ implies $\sum_i x_i(\theta)=1$.
\item \emph{Budgets are respected:} $S_i(\beta;x)\le B_i$.
\item \emph{No unnecessary pacing:} $S_i(\beta;x)<B_i$ implies $\beta_i=0$.
\end{enumerate}
\end{definition}

When $\mu$ is a counting measure this is Definition~1 of \citet{chen2024complexity}. Replacing the price by the winner's own bid gives first-price spend $S^{FP}_i(\beta;x)=\int x_i\,e^{\beta_i}v_i\dd\mu$, and the same four conditions define a \emph{first-price pacing equilibrium} (\fppe) \citep{conitzer2022pacing}. 

\subsection{Fluid Model}
\label{subsec:spend-map}

Motivated by markets with many heterogeneous auction opportunities, we consider a fluid model in which the supply measure $\mu$ is nonatomic \cite{balseiro2015repeated,balseiro2019learning}. This represents a continuum of auction opportunities, with allocations and payments aggregated over the item space. We study finite-item markets separately in \Cref{sec:finite}.

We first identify a compact domain containing all equilibrium multipliers. At any \sppe{}, bidder~$i$'s spending is bounded by $e^{\beta_i}V_i$, since the second-price payment on an item she wins cannot exceed her bid. Thus, a bidder with a multiplier below $B_i/V_i$ cannot exhaust her budget and hence cannot be paced at equilibrium. Defining $\ell_i=\min\{0,\log(B_i/V_i)\}$, every equilibrium lies in the pacing domain $\B=\textstyle\prod_i[\ell_i,0]$.

For every $\beta\in\mathbb R^n$, define the spend map
\[
S_i(\beta)=\int_\Theta
\ind\Bigl\{
b_i(\theta;\beta)>
\max_{k\ne i}b_k(\theta;\beta)
\Bigr\}
\,p(\theta;\beta)\,\dd\mu(\theta),
\] and define $S^{FP}_i(\beta)$ analogously with $p(\theta;\beta)$ replaced by $e^{\beta_i}v_i(\theta)$. These maps depend only on the multiplier profile and capture spending on items for which a bidder is the unique highest bidder. To ensure that they coincide with spending under any allocation satisfying (a) and (b) of \Cref{def:sppe}, we impose the following condition throughout our analysis of the fluid model.

\begin{assumption}[No positive-price ties]
\label{ass:no-ties}
For every $\beta\in\B$, ties for the highest bid at a positive price are $\mu$-null.
\end{assumption}

Under \Cref{ass:no-ties}, for every $\beta\in\B$ and every allocation $x$ satisfying (a) and (b) of \Cref{def:sppe}, we have $S_i(\beta;x)=S_i(\beta)$ and $S^{FP}_i(\beta;x)=S^{FP}_i(\beta)$. Consequently, for $\beta\in\B$, there exists an allocation $x$ such that $(\beta,x)$ is an \sppe{} if and only if, for every bidder $i$,
\[
S_i(\beta)\le B_i,\qquad
S_i(\beta)<B_i\ \Longrightarrow\ \beta_i=0.
\] We therefore identify an \sppe{} in the fluid model with its multiplier profile $\beta$.

The following lemma records basic properties of the spend maps that follow directly from the auction rules. 

\begin{restatable}[Basic properties]
{lemma}{lembasic}
\label{lem:basic}
The spend maps satisfy:
\begin{enumerate}[label=(\roman*)]
\item For every $\beta\in\mathbb R^n$ and bidder $i$, $S_i(\beta)\le e^{\beta_i}V_i$.

\item For every bidder $i$, $S_i$ is nondecreasing in its own log multiplier $\beta_i$.

\item Scaling all bids by the same factor scales spending proportionally: $S(\beta+t\one)=e^tS(\beta)$ for every $\beta\in\mathbb R^n$ and $t\in\mathbb R$.
\end{enumerate}
The corresponding properties hold for $S^{FP}$.
\end{restatable}

To analyze how spending responds to changes in pacing multipliers, we additionally require smoothness of the spend maps.

\begin{assumption}[Smooth spend]\label{ass:smooth}
The spend maps $S$ and $S^{FP}$ are continuously differentiable on an open neighborhood $U\supset\B$.
\end{assumption}

We use \Cref{ass:no-ties,ass:smooth} as high-level regularity conditions in our general analysis of fluid markets. In \Cref{sec:jacobian}, we introduce a distributional regularity assumption that implies both conditions.

\subsection{Best Responses}

Holding rivals' multipliers fixed, bidder $i$'s best response is the largest multiplier consistent with her budget. For $\beta_{-i}\in\B_{-i}$, define
\[
\BR_i(\beta_{-i})
=\max\bigl\{
y\in[\ell_i,0]:
S_i(y,\beta_{-i})\le B_i
\bigr\}.
\] The feasible set is nonempty by \Cref{lem:basic}(i) and closed by continuity. If $S_i(0,\beta_{-i})\le B_i$, then $\BR_i(\beta_{-i})=0$. Otherwise, the best response exhausts the budget: $S_i(\BR_i(\beta_{-i}),\beta_{-i})=B_i.$

We call $\{\beta\in\B:S_i(\beta)=B_i\}$ bidder $i$'s \emph{budget surface}. It is compact by continuity, but may be empty if the bidder's budget never binds.

\begin{restatable}[Equilibria as fixed points]
{lemma}{lemfixed}
\label{lem:fixed}
Under \Cref{ass:smooth}, suppose $\frac{\partial S_i}{\partial\beta_i}(\beta)>0$ for every bidder~$i$ at every $\beta\in\B$ satisfying $S_i(\beta)=B_i$. Then $\beta\in\B$ is an \sppe if and only if it is a fixed point of $\BR=(\BR_1,\ldots,\BR_n)$.
\end{restatable}

\section{Own-Spend Dominance}
\label{sec:thickness}

The fixed-point characterization in \Cref{lem:fixed} suggests best-response iteration as a method for computing \sppe in the fluid model. Establishing convergence of such dynamics, however, requires additional conditions. In general finite-item markets, \sppe need not be unique, and computing one is PPAD-hard \citep{conitzer2022multiplicative,chen2024complexity}. By contrast, first-price pacing equilibria are unique and can be computed efficiently \citep{conitzer2022pacing}. This contrast motivates examining which properties of first-price spending support convergence of pacing dynamics and under what conditions similar properties hold with second-price payments.

The two auction formats differ in how spending responds to changes in pacing multipliers. Under first price, raising a bidder's own multiplier increases payments on items she already wins and may allow her to win additional items. Raising a rival's multiplier, by contrast, can only reduce her spending through lost wins. Under second price, a bidder's own multiplier affects spending only through additional wins, while a rival's higher multiplier has two opposing effects: a \emph{price effect}, which raises payments on retained wins, and a \emph{displacement effect}, which reduces spending through lost wins. We formalize these effects in \Cref{sec:jacobian}.

The different spending responses under first- and second-price payments determine how a bidder's best response reacts to changes in rivals' multipliers. When a bidder is paced, her best response exhausts her budget. The adjustment in her own multiplier following changes in rivals' multipliers therefore depends on how sensitive her spending is to her own multiplier compared with those of her rivals. This motivates the notion of \emph{own-spend dominance}, introduced below. We show that this condition holds under first-price payments whenever spending is positive, but may fail under second-price payments.

To formalize this comparison, let $J(\beta)$ denote the Jacobian of the spend map $S$, with entries
\[
J_{ik}(\beta)
=\frac{\partial S_i}{\partial\beta_k}(\beta),
\quad \beta\in U.
\] The diagonal entries $J_{ii}$ measure the sensitivity of a bidder's spending to her own multiplier, while the off-diagonal entries capture the effects of rivals' multipliers. By \Cref{lem:basic}(ii), $J_{ii}(\beta)\ge 0$. Differentiating the common-scaling property in \Cref{lem:basic}(iii) at $t=0$ yields
\[
J_{ii}(\beta)+\sum_{k\ne i}J_{ik}(\beta)=S_i(\beta),
\] which we refer to as the \emph{scaling identity}.

\subsection{Pointwise Own-Spend Dominance}
\label{subsec:thickness}
Own-spend dominance requires a bidder's spending to be more sensitive to her own log multiplier than to the sum of the absolute effects of changes in rivals' log multipliers.

\begin{definition}[Pointwise own-spend dominance]
At a profile $\beta$, bidder $i$ satisfies \emph{pointwise own-spend dominance} if
\begin{equation}
J_{ii}(\beta)>
\sum_{k\ne i}|J_{ik}(\beta)|.
\label{eq:thickness}
\end{equation}
\end{definition}

This is strict diagonal dominance of row $i$ of the spending Jacobian. If bidder $i$ increases her log multiplier by a small amount $\delta>0$, her spending increases by approximately $\delta J_{ii}$. If instead each rival changes its log multiplier by at most $\delta$ in absolute value, the resulting change in bidder~$i$'s spending is at most $\delta\sum_{k\ne i}|J_{ik}|$ in absolute value, to first order. Own-spend dominance requires the former effect to exceed the latter.

The scaling identity gives an equivalent characterization of own-spend dominance in terms of the positive off-diagonal entries of the spend Jacobian. Let $C_i(\beta):=\sum_{k\ne i}[J_{ik}(\beta)]_+$ denote their sum in row $i$. Using $|x|=2[x]_+-x$, we obtain
\begin{equation}\label{eq:slack}
J_{ii}(\beta)-\sum_{k\ne i}|J_{ik}(\beta)|
=S_i(\beta)-2C_i(\beta).
\end{equation}
Thus, bidder $i$ satisfies pointwise own-spend dominance at $\beta$ if and only if $C_i(\beta)<S_i(\beta)/2$. Although the definition involves all off-diagonal entries of the spend Jacobian, this characterization depends only on their positive parts together with total spending.

\paragraph{Comparison with first-price auctions.}
Under first-price payments, increasing a rival's multiplier cannot increase a bidder's spending. All cross-spending derivatives are therefore nonpositive, making own-spend dominance automatic whenever spending is positive.

\begin{restatable}[First-price diagonal dominance]
{lemma}{lemfpdominance}
\label{lem:first-price-diagonal-dominance}
Suppose $S^{\mathrm{FP}}$ is continuously differentiable on $U$. Then, for every $\beta\in U$ and bidder $i$,
\begin{equation}\label{eq:first-price-dominance}
J_{ik}^{\mathrm{FP}}(\beta)\le0
\quad(k\ne i),
\qquad
J_{ii}^{\mathrm{FP}}(\beta)
-\sum_{k\ne i}|J_{ik}^{\mathrm{FP}}(\beta)|
=S_i^{\mathrm{FP}}(\beta).
\end{equation}
Consequently, the first-price spending Jacobian is strictly diagonally dominant in every row with positive spending.
\end{restatable}
Lemma~\ref{lem:first-price-diagonal-dominance} shows that the aggregate positive cross-spending elasticity under first price is zero wherever spending is positive. By contrast, under second price, positive cross-spending derivatives can cause own-spend dominance to fail, particularly when a bidder wins by a large margin, as the following example illustrates.

\begin{example}[Winning by a margin]
\label{ex:dominance}
Consider two bidders and a profile $\beta\in\B$ at which bidder $1$ exhausts her budget, $S_1(\beta)=B_1$. Suppose bidder $1$ wins every item by a uniform margin: $e^{\beta_1}v_1(\theta) \ge c\,e^{\beta_2}v_2(\theta) \quad\mu\text{-a.e.},$ for some $c>1$, with $V_2>0$. This ranking persists in a neighborhood of $\beta$, so bidder $1$ continues to win every item and pays bidder $2$'s bid. Hence $S_1(\beta')=e^{\beta'_2}V_2$ for all $\beta'$ sufficiently close to $\beta$. It follows that $d_1(\beta)=0$ and $J_{12}(\beta)=S_1(\beta)>0.$ Thus $\Phi_1(\beta)=1$, and pointwise own-spend dominance fails despite bidder $1$ exhausting her budget.
\end{example}

This example illustrates how the absence of \emph{marginal competition} can undermine own-spend dominance: higher rival bids increase payments without displacing the winning bidder. In \Cref{sec:jacobian}, we formalize the price and displacement effects and characterize own-spend dominance in terms of their contributions to the spending derivatives.

\subsection{Uniform Own-Spend Dominance}
\label{subsec:thickness-margins}

Pointwise own-spend dominance compares spending responses at a single profile. To establish convergence of pacing dynamics to an \sppe, we require uniform control of rivals' combined influence relative to the bidder's own spending response across the relevant pacing profiles.

For $\gamma\in(0,1]$, consider the conditions
\begin{equation}\label{eq:uniform-thickness}
J_{ii}(\beta)>0,
\qquad
\sum_{k\ne i}|J_{ik}(\beta)|
\le (1-\gamma)J_{ii}(\beta).
\end{equation}
The parameter $\gamma$ measures the relative diagonal-dominance margin. We distinguish two requirements according to where these inequalities hold:
\begin{itemize}
\item \emph{Uniform own-spend dominance on budget surfaces}, $(D_\gamma)$: \eqref{eq:uniform-thickness} holds for every bidder $i$ at every $\beta\in\B$ satisfying $S_i(\beta)=B_i$.

\item \emph{Uniform own-spend dominance throughout the pacing domain}, $(D_\gamma^+)$: \eqref{eq:uniform-thickness} holds for every bidder~$i$ and every $\beta\in\B$.
\end{itemize}
$(D_\gamma^+)$ implies $(D_\gamma)$. The distinction reflects the domains relevant to different pacing dynamics: best responses are determined on budget surfaces, whereas dual-descent and stochastic pacing dynamics require control throughout the pacing domain.

By \eqref{eq:slack}, the dominance inequality in \eqref{eq:uniform-thickness} can equivalently be expressed in terms of spending as
\begin{equation}\label{eq:uniform-thickness-spend}
S_i(\beta)-2C_i(\beta)
\ge\gamma J_{ii}(\beta).
\end{equation}
In particular, on bidder $i$'s budget surface, \eqref{eq:uniform-thickness-spend} becomes $B_i-2C_i(\beta)\ge\gamma J_{ii}(\beta)$.

The following lemma shows that under \Cref{ass:smooth}, strict pointwise dominance on either domain already guarantees a positive uniform margin.

\begin{restatable}[Existence of a uniform
dominance margin]{lemma}{lemmargin}
\label{lem:thickness-margin}
Under \Cref{ass:smooth}, $(D_\gamma)$ holds for some $\gamma>0$ if and only if pointwise own-spend dominance holds for every bidder $i$ at every $\beta\in\B$ with $S_i(\beta)=B_i$. Likewise, $(D_\gamma^+)$ holds for some $\gamma>0$ if and only if pointwise own-spend dominance holds for every bidder throughout $\B$.
\end{restatable}

Consequently, verifying strict own-spend dominance on the relevant domain suffices to obtain the uniform margin used in the convergence analysis.

\section{The Structure of the Spend Jacobian}
\label{sec:jacobian}

\Cref{sec:thickness} characterized own-spend dominance in terms of the derivatives of the spend maps and introduced the price and displacement effects. We now examine how these effects determine spending responses to small changes in pacing multipliers. We introduce regularity conditions on the distribution of close bidding contests that imply \Cref{ass:no-ties,ass:smooth} and allow us to decompose the spend Jacobian. This decomposition yields a characterization of own-spend dominance in terms of marginal competition.

\subsection{Primitive Regularity Conditions}

Small changes in pacing multipliers can alter allocations when the highest bids on an item are sufficiently close. To characterize the resulting spending derivatives, we study the distribution of log-bid gaps between the highest bidders across items.

For distinct bidders $i$ and $k$, a multiplier profile $\beta\in\mathbb R^n$, and a Borel set $A\subseteq\mathbb R$, define
\[
\nu_{ik}^{\beta}(A) =
\int_\Theta
p(\theta;\beta)
\ind\{\{i,k\}\text{ are the two highest bidders}\}
\ind\{\log b_i-\log b_k\in A\}
\,\dd\mu(\theta).
\] This measure captures the total spending on items where $i$ and $k$ are the two highest bidders and their log-bid difference lies in $A$. In particular, $\nu_{ik}^{\beta}((0,t])$ measures bidder $i$'s spending on items won over bidder $k$ by a log-bid margin of at most $t$. Increasing bidder $k$'s log multiplier by $t$ would displace bidder $i$ on those items with a margin strictly below $t$.

We impose the following regularity conditions on the distribution of these log-bid gaps and on contests involving more than two bidders.

\begin{assumption}[Regularity]
\label{ass:regularity}
Let $U\supset\B$ be an open neighborhood of $\B$.

\begin{enumerate}[label=(R\arabic*)]
\item For every ordered pair $i\ne k$ and every $\beta\in U$, the measure $\nu_{ik}^{\beta}$ has a density $H_{ik}(g;\beta)$ for $g$ in a neighborhood of zero, and $(g,\beta)\mapsto H_{ik}(g;\beta)$ is continuous there.

\item Writing $y_{(1)}\ge y_{(2)}\ge y_{(3)}$ for the three largest log-bids, for every $\beta\in U$ we have $\int_\Theta p(\theta;\beta) \ind\{y_{(1)}-y_{(3)}\le t\} \,\dd\mu(\theta)=o(t)$ as $t\downarrow0,$ and $\mu\bigl( \{y_{(2)}=y_{(3)},\ p>0\} \bigr)=0.$
\end{enumerate}
\end{assumption}

For $n=2$ bidders, we set the third-largest log-bid to $-\infty$, making the conditions involving three bidders vacuous. Zero bids are likewise assigned log-bid $-\infty$, and only positive-price items contribute to $\nu_{ik}^{\beta}$.

The conditions in \Cref{ass:regularity} guarantee that \Cref{ass:no-ties,ass:smooth} hold. Condition (R1) requires a continuous price-weighted density of close contests between the two highest bidders. In particular, $\nu_{ik}^{\beta}((0,t])=tH_{ik}(0;\beta)+o(t),$ so $H_{ik}(0;\beta)$ measures the rate at which bidder $i$'s spending decreases due to displacement when bidder $k$ raises her log multiplier. The continuity requirement ensures that this marginal effect varies continuously with the multiplier profile. Condition (R2) ensures that contests involving three nearly tied bidders have a negligible first-order effect on spending and rules out ties for the second-highest bid at a positive price. Together, these conditions rule out ties for the highest bid at a positive price and allow the spending derivatives to be characterized through pairwise contests, as established in the next subsection.

\subsection{Price and Displacement Effects}
For distinct bidders $i$ and $k$, define
\[
P_{ik}(\beta)
=\int_\Theta p(\theta;\beta)\,
\ind\{i\text{ first},\ k\text{ second}\}
\,\dd\mu(\theta),
\] which measures bidder $i$'s spending on items for which bidder $k$ sets the price, so that $\sum_{k\ne i}P_{ik}(\beta)=S_i(\beta)$. The quantity $P_{ik}(\beta)$ captures the \emph{price effect}: an increase in bidder $k$'s log multiplier raises payments on items that bidder $i$ continues to win.

The \emph{displacement effect}, by contrast, measures the marginal decrease in bidder $i$'s spending from losing narrowly won items when bidder $k$ raises her log multiplier. Under \Cref{ass:regularity}, this effect is determined by the price-weighted density of near ties between the two highest bidders. We therefore define
\begin{equation}\label{eq:D}
D_{ik}(\beta)
:=H_{ik}(0;\beta)
=D_{ki}(\beta).
\end{equation}
The symmetry follows from $H_{ki}(g;\beta)=H_{ik}(-g;\beta)$.

We set $P_{ii}=D_{ii}=0$ and define $L=\diag\Bigl(\textstyle\sum_kD_{ik}\Bigr)-D$ to be the Laplacian of the weighted graph whose edge weights $D_{ik}$ capture the price-weighted density of close contests between the two highest bidders.

\begin{restatable}[Jacobian identity]
{lemma}{lemjacobian}
\label{lem:jacobian}
Under \Cref{ass:regularity}, both \Cref{ass:no-ties,ass:smooth} hold on the same neighborhood~$U$, and
\[
J=P+L,
\qquad
J^{\mathrm{FP}}
=\diag(S^{\mathrm{FP}})+L,
\] with the same matrix $L$. Equivalently, for $k\ne i$,
\[
J_{ik}=P_{ik}-D_{ik},\qquad J_{ii}=\sum_{k\ne i}D_{ik},\qquad J^{FP}_{ik}=-D_{ik},\qquad J^{FP}_{ii}=S^{FP}_i+\sum_{k\ne i}D_{ik}.
\]
\end{restatable}
The Jacobian identity highlights the distinction between the two payment rules. Under second-price auctions, an increase in a rival's multiplier raises payments on retained wins but reduces spending through displacement, whereas an increase in a bidder's own multiplier affects spending only through additional wins. Under first-price auctions, rival multipliers affect spending only through displacement, while an increase in a bidder's own multiplier also raises payments on previously won items.

\begin{remark}[Finite-item markets]
\label{rem:finite}
Suppose $\mu$ is a counting measure on finitely many items. At any profile $\beta$ where the highest and second-highest bidders are uniquely determined on every item, sufficiently small changes in pacing multipliers affect payments but leave allocations unchanged. Consequently, the spend map is locally differentiable with $J=P$ and $D=0$. Displacement occurs only when bid rankings change at discrete thresholds and is therefore not captured by local spending derivatives. In particular, $J_{ii}=0$ for every bidder $i$, so strict own-spend dominance fails at such profiles. In \Cref{sec:finite}, we study finite-item markets through their comparison with fluid markets.
\end{remark}

\paragraph{Marginal competition}
By \Cref{lem:jacobian}, an increase in bidder $k$'s log multiplier raises bidder $i$'s spending at the margin precisely when the price effect exceeds the displacement effect, $P_{ik}>D_{ik}$. The positive part $[P_{ik}-D_{ik}]_+$ measures the remaining price effect on bidder $i$'s spending after offsetting by displacement, whereas $\min\{P_{ik},D_{ik}\}$ measures the portion offset by displacement, capped at the spending for which bidder $k$ sets the price.

\begin{definition}[Marginal competition]
\label{def:marginal-competition}
At a multiplier profile $\beta$, the \emph{marginal competition} faced by bidder $i$ is
\[
M_i(\beta)
:=\sum_{k\ne i}
\min\{P_{ik}(\beta),D_{ik}(\beta)\}.
\] Bidder $i$ faces \emph{sufficient marginal competition} at $\beta$ if $M_i(\beta)>\frac12 S_i(\beta)$.
\end{definition}
Since $\sum_{k\ne i}P_{ik}=S_i$, we have $M_i =\sum_{k\ne i}\bigl(P_{ik}-[P_{ik}-D_{ik}]_+\bigr)=S_i-C_i.$ Together with \eqref{eq:slack}, this gives $J_{ii}-\sum_{k\ne i}|J_{ik}|=2M_i-S_i.$ Hence sufficient marginal competition is equivalent to pointwise own-spend dominance, or, equivalently,
\begin{equation}\label{eq:PD-thickness}
\sum_{k\ne i}
[P_{ik}(\beta)-D_{ik}(\beta)]_+
<\frac12 S_i(\beta).
\end{equation}
Similarly, substituting $J_{ii}=\sum_{k\ne i}D_{ik}$ into \eqref{eq:uniform-thickness}, the uniform own-spend dominance condition becomes
\begin{equation}
    \sum_{k\ne i}D_{ik}(\beta)>0,
    \qquad
    S_i(\beta)
    -2\sum_{k\ne i}[P_{ik}(\beta)-D_{ik}(\beta)]_+
    \geq\gamma\sum_{k\ne i}D_{ik}(\beta).
    \label{eq:PD-uniform-thickness}
\end{equation}

Thus, sufficient marginal competition requires displacement to offset more than half of the bidder's total price effect, counting displacement separately against each rival and only up to the price effect generated by that rival. A bidder may therefore face rivals that raise its prices without displacing it at the margin, provided the aggregate unoffset price effects account for less than half of its spending.

\subsection{Economic Interpretation of Marginal Competition}
\label{subsec:interpret}

Marginal competition has a simple economic interpretation in terms of two properties of each competitor: how much of a bidder's spending it determines, and how strongly it displaces the bidder when bidding more aggressively. For bidder $i$ with $S_i>0$, the condition $M_i>\frac12 S_i$ is equivalent to
\begin{equation}\label{eq:thickness-interpretation}
\sum_{k\ne i:P_{ik}>0}
\frac{P_{ik}}{S_i}
\min\left\{1,\frac{D_{ik}}{P_{ik}}\right\}
>\frac12.
\end{equation}
Here, $P_{ik}/S_i$ is the share of bidder $i$'s spending priced by rival $k$, while $D_{ik}/P_{ik}$ measures the displacement effect relative to that rival's price effect -- that is, bidder $i$'s relative sensitivity to competitive displacement. Thus, the bidder faces sufficient marginal competition precisely when the spending-weighted average of these ratios, capped at one, exceeds one half. The cap reflects that each rival's displacement is credited only up to the spending it prices.

In particular, $D_{ik}/P_{ik}>1/2$ for every price-setting rival is sufficient, but not necessary. Rivals with weak displacement effects can be offset by others with stronger effects, weighted by their shares of the bidder's spending. Marginal competition therefore need not be strong against every rival or on every item type. Rather, it requires sufficient displacement in aggregate and rules out markets in which rivals can raise prices on most of a bidder's spending without displacing it at the margin, as illustrated in \Cref{ex:dominance}.

This characterization motivates the search for \emph{economically interpretable sufficient conditions for marginal competition}. The relevant conditions depend on the market structure and on how the displacement-to-price ratios can be evaluated or bounded. The following sections develop such conditions based on winning probabilities, properties of bid distributions, and responsiveness to finite changes in competing bids.
\section{Own-Spend Dominance Implies Contraction}\label{sec:contraction}

\Cref{sec:jacobian} characterizes own-spend dominance in terms of marginal competition. We now show that when this condition holds uniformly over the relevant pacing profiles, convergence to a unique \sppe is guaranteed under three adjustment dynamics: best responses, projected dual descent, and per-auction stochastic pacing.

The domain on which own-spend dominance must hold differs across these dynamics. Best-response contraction requires it only for \((D_\gamma)\), since paced best responses exhaust the bidder's budget, and their sensitivity to rivals' multipliers is therefore determined by spending derivatives on budget surfaces. By contrast, spend-based dynamics adjust multipliers at profiles where budgets need not be exhausted, so their convergence guarantees require it on \((D_\gamma^+)\).

\subsection{Best-Response Dynamics}
Own-spend dominance limits the sensitivity of best responses to changes in rivals' multipliers. At an interior best response, the bidder exhausts her budget, so changes in rivals' multipliers that affect her spending require an adjustment to her own multiplier. Under own-spend dominance, this adjustment is smaller, to first order, than the largest absolute change in any rival's log multiplier.

To formalize this, suppose bidder $i$ has an interior best response, $\BR_i(\beta_{-i})\in(\ell_i,0)$. Since her budget is exhausted, $S_i(\BR_i(\beta_{-i}),\beta_{-i})=B_i.$ If $J_{ii}>0$ at the corresponding profile, the implicit function theorem implies that $\BR_i$ is locally differentiable, with
\begin{equation}\label{eq:dbr}
\frac{\partial \BR_i}{\partial\beta_k}
=-\frac{J_{ik}}{J_{ii}},
\qquad k\ne i,
\end{equation}
where the derivatives on the right are evaluated at $(\BR_i(\beta_{-i}),\beta_{-i})$.

Thus, if each rival's log multiplier changes by at most $\delta$, the resulting change in bidder $i$'s best response is bounded, to first order, by $\delta\sum_{k\ne i}|J_{ik}|/J_{ii}$. Pointwise own-spend dominance makes this ratio strictly less than one, while $(D_\gamma)$ bounds it uniformly by $1-\gamma$. The following theorem extends this local bound to a global contraction.

\begin{restatable}[Best-response contraction]{theorem}{thmbr}\label{thm:br}
Under \Cref{ass:smooth} and $(D_\gamma)$, $\BR:\B\to\B$ is a $(1-\gamma)$-contraction in $\norm{\cdot}_\infty$. Consequently, there is a unique \sppe $\beta^*$, synchronous best-response dynamics $\beta^{t+1}=\BR(\beta^t)$ satisfy
\[
\norm{\beta^t-\beta^*}_\infty\le(1-\gamma)^t\norm{\beta^0-\beta^*}_\infty,
\] and totally asynchronous best-response dynamics converge to $\beta^*$.
\end{restatable}
The proof extends the local sensitivity bound in \eqref{eq:dbr} to boundary best responses, where budget exhaustion need not hold.

\begin{remark}[Monotone best responses]
\label{rem:monotone}
Suppose, in addition, that $J_{ik}(\beta)\le0$ for every $i$, $k\ne i$, and $\beta\in\B$ with $S_i(\beta)=B_i$. Under \Cref{ass:regularity}, this is equivalent to $P_{ik}\le D_{ik}$: each rival's displacement effect is at least as large as its price effect. By \eqref{eq:dbr}, $\partial\BR_i/\partial\beta_k\ge0$ at interior best responses. Together with the continuity of $\BR_i$ at the pacing boundary (see proof of \Cref{thm:br}), this implies that $\BR$ is isotone. Consequently, best-response dynamics initialized at $\beta^0=0$ satisfy $\beta^1=\BR(0)\le\beta^0$ and $\beta^{t+1}\le\beta^t$ for all $t$, and converge monotonically to $\beta^*$.
\end{remark}

In first-price auctions, uniform own-spend dominance holds automatically. Let $\BR^{FP}$ denote the best response function defined using first-price spending $S^{FP}$.

\begin{restatable}[First-price best responses]{corollary}{corfpbr}
\label{cor:fp-br}
Under \Cref{ass:smooth}, first-price spending satisfies $(D_\gamma)$ for some $\gamma>0$. In particular, one may take
\[
\gamma^{FP}=
\min\left\{
1,\,
\inf_{\substack{i,\ \beta\in\B:\\
S_i^{FP}(\beta)=B_i}}
\frac{B_i}{J_{ii}^{FP}(\beta)}
\right\}>0,
\] where the infimum over an empty set is understood as $+\infty$. Consequently, in the fluid model, $\BR^{FP}:\B\to\B$ is isotone and a $(1-\gamma^{FP})$-contraction in $\norm{\cdot}_\infty$. There is a unique \fppe, and synchronous first-price best-response dynamics converge to it geometrically from every initial profile in $\B$.
\end{restatable}

Thus, unlike under second-price payments, first-price best-response contraction requires no additional conditions on competition in the fluid model. Uniqueness of \fppe also follows from its convex-program characterization \citep{conitzer2022pacing}; \Cref{cor:fp-br} establishes it through best-response contraction.

\subsection{Spend-Based Dynamics}

We next consider dynamics that adjust pacing multipliers according to deviations of spending from budget targets. A bidder's multiplier decreases when spending exceeds her budget and increases when spending falls below it, subject to the pacing constraints. We study two such dynamics: projected dual descent, which updates multipliers using expected spending, and per-auction stochastic pacing, which uses realized payments. Theorems~\ref{thm:dd} and~\ref{thm:sa} show that these dynamics converge to the unique \sppe under uniform own-spend dominance.

\begin{restatable}[Projected dual descent]{theorem}{thmdd}
\label{thm:dd}
Under \Cref{ass:smooth} and $(D^+_\gamma)$, let $J_{\max}=\max_{\beta\in\B}\max_i J_{ii}(\beta)$, and $J_{\min}=\min_{\beta\in\B}\min_i J_{ii}(\beta)>0$. For any step sizes $0<\eta_i\le 1/J_{\max}$, the map
\[
T(\beta)=\Pi_\B\bigl[\beta-\eta(S(\beta)-B)\bigr]
\] is a contraction in $\norm{\cdot}_\infty$ with contraction factor at most $1-\gamma J_{\min}\min_i\eta_i$. Its unique fixed point is the unique \sppe{} $\beta^*$.
\end{restatable}

The contraction follows directly from uniform own-spend dominance. The Jacobian of the unprojected update is $I-\diag(\eta)J$, whose absolute row sums satisfy $1-\eta_i J_{ii}+\eta_i\sum_{k\ne i}|J_{ik}| \le 1-\eta_i\gamma J_{ii}.$ Since projection onto $\B$ is nonexpansive, the same bound applies to the projected update.

Projected dual descent requires evaluating the spend map at each iteration. Alternatively, updates can be based on payments observed in individual auctions, which provide unbiased estimates of expected spending. The resulting stochastic dynamics use decreasing step sizes to control sampling noise.

\begin{restatable}[Per-auction stochastic pacing]{theorem}{thmsa}
\label{thm:sa}
Let $\mu$ be a probability measure, with $\int p(\theta;0)^2\dd\mu(\theta)<~\infty,$ and interpret $B_i$ as bidder $i$'s budget per auction. Suppose items $\theta_t\sim\mu$ arrive i.i.d., and let $z_i^t$ denote bidder $i$'s realized payment in auction $t$, under any tie-breaking rule. Consider the updates
\[
\beta^{t+1}
=\Pi_\B\bigl[\beta^t-\epsilon_t\,c(z^t-B)\bigr],
\] where $c_i>0$ and the step sizes satisfy $\sum_t\epsilon_t=\infty,\, \sum_t\epsilon_t^2<\infty$. Under \Cref{ass:smooth} and $(D^+_\gamma)$, $\beta^t\to\beta^*$ almost surely.
\end{restatable}

\section{Logit Markets}
\label{sec:logit}
In \Cref{sec:contraction}, we established that uniform own-spend dominance guarantees uniqueness of \sppe and convergence of best-response and spend-based dynamics. We now derive sufficient conditions for this property from the underlying market structure. 

In internet advertising auctions, a bidder's expected value for an impression depends on her value per click and the impression's click-through probability \citep{varian2007position, edelman2007internet}. Advertising platforms estimate these probabilities using machine-learning models \citep{richardson2007predicting,mcmahan2013ad,he2014practical}. Variation in click-through probabilities across impressions generates heterogeneity in bidders' valuations, even within the same item type \citep{perlich2012bid,balseiro2015repeated}. We model this heterogeneity through independent Gumbel-distributed variation in bidders' log-valuations, which yields logit winning probabilities\footnote{ Related work has modeled bidder values in sponsored-search auctions using lognormal distributions \citep{ostrovsky2023reserve}.} \citep{mcfadden1972conditional}. This setup provides a tractable setting for analyzing own-spend dominance: the price and displacement effects characterized in \Cref{sec:jacobian} admit closed-form expressions, allowing us to derive sufficient conditions for own-spend dominance in terms of bidders' win shares.

\paragraph{Market Model.}
We consider a market with item types $j=1,\ldots,Q$, each with mass $\mu_j>0$. On an item of type $j$, bidder $i$'s log-value is $\log v_{ij}=a_{ij}+\sigma\varepsilon_{ij},$ where the $\varepsilon_{ij}$ are independent standard Gumbel random variables and $\sigma\in(0,1)$. The parameters $a_{ij}$ capture systematic differences in bidders' valuations, while $\varepsilon_{ij}$, drawn independently across impressions, captures impression-specific variation whose magnitude is controlled by $\sigma$. At a pacing profile $\beta$, bidder $i$'s log-bid on type $j$ is $\beta_i+a_{ij}+\sigma\varepsilon_{ij}$. Define $q_{ij}=e^{(\beta_i+a_{ij})/\sigma},$ $\bar q_j=\sum_k q_{kj},$ and $q_{-i,j}=\bar q_j-q_{ij}.$ Bidder $i$ wins an item of type $j$ with probability $w_{ij}={q_{ij}}/{\bar q_j}$. We refer to $w_{ij}$ as bidder $i$'s \emph{win share} on type $j$.

We call $q_{ij}$ bidder $i$'s \emph{logit weight} on type $j$. These weights add across bidders in a useful sense. The highest bid among any group of bidders has the same distribution as the bid of a single bidder whose weight is the sum of their weights. Thus bidder $i$ faces a single effective rival of weight $q_{-i,j}$ when determining whether she wins and how much she pays. A bidder of weight $q$ has mean bid $m(q)=\Gamma(1-\sigma)q^\sigma$, where $\Gamma(1-\sigma)$ is a scale factor determined by the dispersion parameter $\sigma$.\footnote{The Gamma function is defined by $\Gamma(z)=\int_0^\infty t^{z-1}e^{-t}\,dt$ for $z>0$.}

\begin{restatable}[Logit closed forms]{lemma}{lemlogit}
\label{lem:logit}
The logit market satisfies \Cref{ass:regularity} with $U=\R^n$. Spending and the price and displacement effects aggregate over types:
\[
S_i=\sum_j\mu_jS_i^{(j)},\qquad
P_{ik}=\sum_j\mu_jP_{ik}^{(j)},\qquad
D_{ik}=\sum_j\mu_jD_{ik}^{(j)},
\] where
\begin{align*}
S_i^{(j)}
&=\Gamma(1-\sigma)
  \bigl[q_{-i,j}^{\,\sigma}
  -q_{-i,j}\bar q_j^{\,\sigma-1}\bigr],\\
P_{ik}^{(j)}
&=\Gamma(1-\sigma)\,q_{kj}
  \bigl[q_{-i,j}^{\,\sigma-1}
  -\bar q_j^{\,\sigma-1}\bigr],\\
D_{ik}^{(j)}
&=\frac{\Gamma(2-\sigma)}{\sigma}
  q_{ij}q_{kj}\bar q_j^{\,\sigma-2}.
\end{align*}
Moreover, for every rival $k\ne i$, the price effect to displacement effect ratio satisfies
\begin{equation*}
\frac{P_{ik}^{(j)}}{D_{ik}^{(j)}}
=\rho(\sigma,w_{ij}),
\qquad
\rho(\sigma,w)
=\frac{\sigma}{1-\sigma}
\frac{(1-w)^{-(1-\sigma)}-1}{w}.
\end{equation*}
\end{restatable}
The spending formula can be understood as follows. Start with the effective rival's mean bid, $m(q_{-i,j})$. Bidder $i$ pays this rival's realized bid whenever she wins and pays zero whenever she loses. On the impressions she loses, the rival's bid is the highest bid overall. In the logit model the highest bid is independent of the winner's identity, so its mean on those impressions is $m(\bar q_j)$. Subtracting that mean weighted by the probability of a loss gives
\[
S_i^{(j)}=m(q_{-i,j})-(1-w_{ij})\,m(\bar q_j).
\]

To recover the price effect of an individual rival $k$, observe that when $i$ wins, $k$ sets the price with probability $q_{kj}/q_{-i,j}$, independently of the price. Thus $k$ sets the price on a fraction $q_{kj}/q_{-i,j}$ of $i$'s expected spending on type $j$. Raising $k$'s log multiplier increases those payments proportionally, giving $P_{ik}^{(j)}=(q_{kj}/q_{-i,j})S_i^{(j)}$.

The displacement effect, by contrast, depends on the price-weighted density of impressions where bidders $i$ and $k$ are nearly tied for the highest bid. In the logit model, the price-weighted frequency of such close contests is proportional to the product $w_{ij}w_{kj}$, as can be seen by writing $D_{ik}^{(j)}=\frac{\Gamma(2-\sigma)}{\sigma}\bar q_j^\sigma w_{ij}w_{kj}.$ Thus, displacement is limited when either bidder has a small win share. 

Moreover, both $P_{ik}^{(j)}$ and $D_{ik}^{(j)}$ are proportional to $q_{kj}$: a rival that accounts for twice as much of bidder $i$'s spending as another rival on type $j$ also contributes twice as much to her marginal displacement. Consequently, the ratio $\rho(\sigma,w_{ij})$ of price to displacement effects is independent of the rival $k$ and depends only on bidder $i$'s win share $w_{ij}$. The following lemma establishes how this ratio varies with $w_{ij}$.

\begin{restatable}[Price effects and win shares]
{lemma}{lemrho}
\label{lem:rho}
For $\sigma\in(0,1)$, the function $\rho(\sigma,w)$, extended to $w=0$ by continuity, is continuous and strictly increasing on $[0,1)$, with $\rho(\sigma,0)=\sigma$ and $\lim_{w\uparrow 1}\rho(\sigma,w)=\infty.$ In particular, there are unique $w_1(\sigma)<w_2(\sigma)$ in $(0,1)$ satisfying $\rho(\sigma,w_1(\sigma))=1,$ and $\rho(\sigma,w_2(\sigma))=~2.$
\end{restatable}

By \Cref{lem:rho}, the ratio increases with bidder $i$'s win share. When a bidder dominates an item type, winning nearly every impression of that type, higher rival bids primarily raise the prices she pays rather than displace her. Conversely, lower win shares correspond to greater displacement relative to the price effect. In particular, the thresholds $w_1(\sigma)$ and $w_2(\sigma)$ identify the win shares at which displacement offsets all or half of the price effect, respectively. Together with \eqref{eq:thickness-interpretation}, this relationship yields sufficient conditions for own-spend dominance in terms of win shares and spending shares.

\subsection{A Sufficient Spending-Weighted Win-Share Condition}\label{sec:win-share-condition}

Although price and displacement effects are characterized separately on each item type, own-spend dominance depends on their aggregate contributions across types. Own-spend dominance, however, depends on these effects aggregated across item types. For $S_i>0$, let $s_{ij}:=\mu_j S_i^{(j)}/S_i$ denote bidder $i$'s spending share on type $j$, and define
\[
e_{ij}:=\left(1-\frac{1}{\rho(\sigma,w_{ij})}\right)_+
=\frac{(P_{ik}^{(j)}-D_{ik}^{(j)})_+}{P_{ik}^{(j)}},
\qquad \text{for } k\ne i \text{ with } P_{ik}^{(j)}>0.
\] Thus, $e_{ij}$ measures the fraction of the price effect on type $j$ not offset by displacement. Since all terms in this expression depend on the rival $k$ only through the common factor $q_{kj}$, this factor cancels in the ratio, making $e_{ij}$ independent of $k$.

The following proposition uses these fractions and spending shares to establish a sufficient condition for own-spend dominance.

\begin{restatable}[Spending-weighted win-share test]
{proposition}{propdominance}
\label{prop:dominance}
In the logit market, at any profile $\beta$ with $S_i(\beta)>0$, $C_i\le S_i\sum_j s_{ij}e_{ij},$ and consequently $J_{ii}-\sum_{k\ne i}|J_{ik}|\ge S_i\left(1-2\sum_j s_{ij}e_{ij}\right).$ In particular, bidder $i$ satisfies pointwise own-spend dominance at $\beta$ whenever $\sum_j s_{ij}e_{ij}<\frac12$.
\end{restatable}

\Cref{prop:dominance} shows that own-spend dominance holds when the spending-weighted average of the fractions of price effects not offset by displacement is below one half. Because each type's contribution is weighted by its spending share, high win shares on some types need not preclude own-spend dominance, provided displacement sufficiently offsets price effects on types accounting for most of the bidder's spending. The condition is sufficient but not necessary, since the bound ignores cancellations between positive and negative effects across item types.

The thresholds in \Cref{lem:rho} allow us to bound $e_{ij}$ using bidder $i$'s win share. On item types where $w_{ij}\le w_1(\sigma)$, displacement fully offsets the price effect, so $e_{ij}=0$. Where $w_{ij}<w_2(\sigma)$, displacement offsets more than half of the price effect, so $e_{ij}<1/2$. These bounds yield simpler sufficient conditions for own-spend dominance in terms of win shares and spending shares. We emphasize that these are simple sufficient conditions; one can derive stronger conditions as well.

\begin{restatable}[Sufficient win-share conditions]
{corollary}{corlogithalf}
\label{cor:logit-half}
In the logit market, at a profile $\beta$ with $S_i(\beta)>0$, each of the following conditions implies that bidder $i$ satisfies pointwise own-spend dominance:

\begin{enumerate}[label=(\alph*)]

\item $w_{ij}<w_2(\sigma)$ on every item type $j$ with $s_{ij}>0$;
\item more than half of bidder $i$'s spending is on item types with $w_{ij}\le w_1(\sigma)$.
\end{enumerate}
\end{restatable}

Condition (a) requires sufficiently low win shares on every item type where bidder $i$ spends, so that displacement offsets more than half of the price effect. Condition (b) instead requires displacement to fully offset the price effect on types accounting for more than half of her spending, allowing arbitrarily high win shares on the remaining types.

\begin{remark}[Relative bid levels and win shares]
The win-share conditions can also be related to competitors' relative bid levels. Bidder $k$'s mean paced bid on type $j$ is $\E[e^{\beta_k}v_{kj}] =\Gamma(1-\sigma)e^{\beta_k+a_{kj}}$. A higher mean paced bid for rival $k$ relative to bidder $i$'s reduces her win share, thereby increasing the displacement effect relative to the price effect. To formalize this, fix $c>0$ and let $\bar w_c=1/(1+c^{1/\sigma})$. If some rival's mean paced bid on type $j$ is at least $c$ times bidder $i$'s, then $w_{ij}\le\bar w_c$, and hence $e_{ij}\le(1-\frac{1}{\rho(\sigma,\bar w_c)})_+.$ Bounding $e_{ij}$ by one on the remaining types, \Cref{prop:dominance} implies own-spend dominance if more than a fraction $\max\{1,\rho(\sigma,\bar w_c)\}/2$ of bidder $i$'s spending occurs on types with such a rival, provided this threshold is less than one.
\end{remark}

\begin{example}[Win-share conditions]
For $\sigma=1/2$, the win-share thresholds take the simple forms $w_1(1/2)=(\sqrt{5}-1)/2\approx 61.8\%$ and $w_2(1/2)=\sqrt{3}/2\approx 86.6\%$. Thus, own-spend dominance holds if bidder $i$'s win share is below $86.6\%$ on every item type where she spends, or if more than half of her spending occurs on types with win shares at most $61.8\%$.

Alternatively, a rival whose mean paced bid on type $j$ is at least half of bidder $i$'s ensures that bidder $i$ wins at most $80\%$ of impressions of that type. Own-spend dominance then holds if types with at least one such rival account for more than $77.3\%$ of bidder $i$'s total spending.
\end{example}

\subsection{Uniform Win-Share Conditions}
\Cref{sec:win-share-condition} establishes sufficient conditions for own-spend dominance at a given pacing profile. For the convergence guarantees, however, dominance must hold uniformly across the relevant profiles. Verifying the spending-weighted condition at every pacing profile in the domain can be difficult, since both win shares and spending shares depend on the pacing multipliers.

A simpler, though more restrictive, approach is to bound win shares uniformly across bidders, item types, and pacing profiles. By \Cref{lem:rho}, a uniform upper bound strictly below $w_2(\sigma)$ ensures that displacement offsets more than half of the price effect on every item type, independently of which rival sets the price. This guarantees uniform own-spend dominance, as formalized in the following theorem.

\begin{restatable}[Uniform own-spend dominance]
{theorem}{thmlogit}
\label{thm:logit}
In the logit market, let $\omega=\max_{i,j}\max_{\beta\in\B}w_{ij}(\beta).$ If $\rho(\sigma,\omega)<2$, equivalently $\omega<w_2(\sigma)$, then $(D_\gamma^+)$ holds with $\gamma=\min\{\sigma,\,2-\rho(\sigma,\omega)\}.$ Consequently, the \sppe is unique, best responses contract with factor $\max\{1-\sigma,\,\rho(\sigma,\omega)-1\}$, projected dual descent is a contraction, and per-auction stochastic pacing converges almost surely, as established in \Cref{thm:br,thm:dd,thm:sa}.

If moreover $\rho(\sigma,\omega)\le1$, then $J_{ik}\le0$ for all $i\ne k$ throughout $\B$, so best responses are also isotone (\Cref{rem:monotone}).
\end{restatable}

\Cref{thm:logit} gives a sufficient condition for convergence of pacing dynamics in terms of the maximum win share attained over all bidders, item types, and pacing profiles in $\B$. The condition is stronger than the spending-weighted criterion of \Cref{prop:dominance}: it requires sufficient competition on every item type, even those accounting for little of a bidder's spending. Its advantage is that convergence can be established from a single bound on the maximum win share, without controlling spending shares across pacing profiles.

We next express this maximum in terms of the log-value parameters and pacing bounds. Expressing bidder $i$'s win share in terms of rivals' relative logit weights gives
\[
w_{ij}(\beta)
=\left(1+\sum_{k\ne i}\frac{q_{kj}}{q_{ij}}\right)^{-1},
\qquad
\frac{q_{kj}}{q_{ij}}
=e^{(\beta_k+a_{kj}-\beta_i-a_{ij})/\sigma}.
\] Each relative weight is minimized when bidder $i$ is unpaced and rival $k$ is paced to its lower bound $\ell_k$. Since these minima are attained simultaneously, for every $\beta\in\B$,
\begin{equation}\label{eq:share-bound}
w_{ij}(\beta)
\le
\left[
1+\sum_{k\ne i}
e^{(a_{kj}+\ell_k-a_{ij})/\sigma}
\right]^{-1},
\end{equation}
with equality when $\beta_i=0$ and $\beta_k=\ell_k$ for all $k\ne i$. 
Hence, maximizing the right-hand side of \eqref{eq:share-bound} over $i$ and $j$ yields $\omega$. It follows that $\omega<w_2(\sigma)$ if and only if
\begin{equation}\label{eq:large-cond}
    \sum_{k\ne i}e^{(a_{kj}+\ell_k-a_{ij})/\sigma}> \frac{1}{w_2(\sigma)}-1
    \qquad\text{for all }i,j.
\end{equation}

Each term in \eqref{eq:large-cond} is rival $k$'s relative logit weight at the profile where bidder $i$'s win share is maximized. The condition therefore requires a lower bound on the aggregate relative weight of rivals, without imposing such a bound on any individual rival: several weaker rivals can together limit bidder $i$’s win share as effectively as a single strong one. The following proposition provides a sufficient condition in terms of the number of rivals whose mean paced bids, even at their lower bounds, remain at least a fixed fraction of bidder $i$'s unpaced mean bid.

\begin{restatable}[Enough comparable rivals]
{proposition}{proplarge}
\label{prop:large}
In the logit market, fix $W\ge0$, and call rival $k$ \emph{comparable} to bidder $i$ on type $j$ if $a_{kj}+\ell_k\ge a_{ij}-W$. If every bidder has at least $N$ comparable rivals on every item type, then
\[
\omega\le(1+Ne^{-W/\sigma})^{-1}.
\] Consequently, \Cref{thm:logit} applies whenever
\[
N>e^{W/\sigma}
\left(\frac{1}{w_2(\sigma)}-1\right).
\] As $N\to\infty$ with $W$ fixed, the best-response contraction-factor bound tends to $1-\sigma$.
\end{restatable}

\Cref{prop:large} shows that the number of rivals can compensate for differences in bid levels: since $e^{(a_{kj}+\ell_k-a_{ij})/\sigma} \ge e^{-W/\sigma},$ each comparable rival contributes at least $e^{-W/\sigma}$ to the sum in \eqref{eq:large-cond}, so $N$ such rivals contribute at least $Ne^{-W/\sigma}$, which gives the stated condition on $N$. As $W$ increases, this minimum contribution decreases exponentially in $W/\sigma$, so more comparable rivals are required. Conversely, for fixed $W$, increasing the number of comparable rivals on every item type drives $\omega$ toward zero. By \Cref{lem:rho}, the ratio of price to displacement effects then approaches $\sigma<1$, and the contraction-factor bound approaches $1-\sigma$. Thus, the large-market condition in \Cref{prop:large} requires sufficiently many comparable rivals for each bidder on each item type, rather than merely a large total number of bidders.

\section{From Fluid Markets to Finite Instances}\label{sec:finite}

Unlike in a fluid market, the spend map of a finite market can jump when changes in pacing multipliers alter the identity of an item's winning bidder. Between such changes, allocations remain fixed, so local spending derivatives capture only the price effect, not displacement (\Cref{rem:finite}). Consequently, the derivative-based contraction arguments of \Cref{sec:contraction} do not apply directly.

We instead compare the finite spend map with that of a fluid market satisfying own-spend dominance. If the two spend maps are \emph{uniformly close}, and fluid spending increases sufficiently with a bidder's own multiplier near her budget, then their best responses are also \emph{close}. Finite best-response dynamics therefore inherit the fluid contraction up to an approximation error. Finally, rescaling all pacing multipliers by a common factor preserves bid rankings while reducing payments proportionally, yielding an allocation that satisfies the budget constraints exactly. We measure the resulting approximation using the notion of approximate \sppe introduced by~\citet{chen2024complexity}.

\begin{definition}[Approximate \sppe, \citealp{chen2024complexity}]\label{def:approx}
For $\delta,\phi\ge0$, a pair $(\beta,x)$ is a \emph{$(\delta,\phi)$-approximate \sppe} if conditions (b) and (c) of \Cref{def:sppe} hold, and conditions (a) and (d) are relaxed to
\begin{align*}
\text{(a$'$)}\qquad x_i(\theta)>0&\ \Longrightarrow\ b_i(\theta;\beta)\ge(1-\delta)h(\theta;\beta),\\
\text{(d$'$)}\qquad S_i(\beta;x)<(1-\phi)B_i&\ \Longrightarrow\ e^{\beta_i}\ge1-\phi.
\end{align*}
\end{definition}

Thus $\delta$ controls how close a winning bid must be to the highest bid, and $\phi$ controls how much an underspending bidder may be paced.

\subsection{Spend-Map Approximation}
\label{subsec:spend-approx}

Throughout this section, we let $\mu$ be the probability measure describing the fluid market, with per-item budgets $B_i$, spend map $S$, and \sppe $\beta^*$. A finite instance $I$ consists of $m$ items $\theta_1,\ldots,\theta_m$, with values $v_i(\theta_1),\ldots,v_i(\theta_m)$ and budgets $mB_i$.  We normalize finite spending by $m$, so that bidder $i$'s budget per item is $B_i$ in both the finite and fluid models.

Unlike in the fluid model, ties for the highest bid can occur in a finite instance, so spending may depend on how tied items are allocated. For each bidder $i$, let $\hat S_i^{>}(\beta)$ denote her spending per item when only items on which she is the unique highest bidder are counted. Let $\hat S_i^{\ge}(\beta)$ also include items on which she ties for the highest bid. Under any allocation to highest bidders, her spending per item lies between these two quantities. The upper bounds need not correspond to a single allocation, since a tied item cannot be fully allocated to every tied bidder.

We define the finite best-response map by
\[
\widehat{\BR}_i(\beta_{-i})=
\sup\{x\le 0:\hat S_i^{>}(x,\beta_{-i})\le B_i\}.
\] Holding rivals' multipliers fixed, each item that bidder $i$ can win has a threshold at which her bid equals the highest rival bid. Above this threshold, she wins the item and pays the highest rival bid, independently of her own multiplier. Thus, $\hat S_i^{>}$ is a nondecreasing step function of her multiplier, and $\widehat{\BR}_i$ can be computed by sorting these thresholds and accumulating payments up to the budget. 

To compare finite and fluid best responses, we require the finite spend maps to approximate fluid spending $S$ uniformly over the relevant pacing profiles. Controlling both $\hat S_i^{>}$ and $\hat S_i^{\ge}$ ensures that the approximation holds under any allocation of tied items.

\begin{definition}[Closeness]\label{def:close}
A finite instance $I$ is \emph{$\varepsilon$-close} to $\mu$ on a set $\mathcal C\subseteq\R^n_{\le0}$ if $\abs{\hat S_i^{>}(\beta)-S_i(\beta)}\le\varepsilon$ and $\abs{\hat S^{\ge}_i(\beta)-S_i(\beta)}\le\varepsilon$ for every bidder $i$ and every $\beta\in\mathcal C$.
\end{definition}

When items are drawn independently from $\mu$, finite spending per item is a sample average whose expectation equals fluid spending at each fixed pacing profile. The main challenge is to bound the approximation error uniformly over all profiles, since changes in pacing multipliers can alter the winning bidder and make finite spending discontinuous.
The underlying winning and payment events, however, have a simple structure. A bidder wins an item when her paced bid exceeds every rival's bid, and she then pays the highest rival bid. Consequently, the event that she wins an item and pays more than a given amount is determined by these bid comparisons, together with whether at least one rival's bid exceeds that amount. The following lemma uses this structure to establish a uniform approximation bound.

\begin{restatable}[Random instances are close]{lemma}{lemvc}\label{lem:vc}
Suppose \Cref{ass:smooth} holds, $0\le v_i\le\bar v$ for every $i$, and the $m$ items are drawn i.i.d.\ from $\mu$. Then for every $\varrho\in(0,1]$, with probability at least $1-\varrho$ the instance is $\varepsilon_m$-close to $\mu$ on $U\cap\R^n_{\le0}$, where
\[
\varepsilon_m=\bar v\sqrt{\frac{32\bigl(2n\log(m+1)+\log(16n/\varrho)\bigr)}{m}} .
\]
\end{restatable}

Since payments are bounded by $\bar v$, standard concentration bounds give an error of order $\bar v/\sqrt{m}$ at each fixed pacing profile.
Extending this bound uniformly over profiles relies on the structure of winning and payment events: as the pacing multipliers vary, each comparison defines a nested family containing at most $m+1$ distinct subsets of the sampled items. Consequently, the number of possible event patterns is polynomial in $m$ for fixed $n$. A Vapnik--Chervonenkis inequality then yields the uniform spending-error bound, including at profiles with bid ties, under either tie convention.

\subsection{Approximate Contraction}
\label{subsec:transfer}

Although fluid best responses lie in $\B$, finite best responses need not. At the fluid lower bound $\ell_i$, bidder $i$'s spending is at most her budget, but the finite spend map may exceed the fluid spend map by a small amount. A finite best response may therefore fall below $\ell_i$. To account for this undershooting, the discrete approximation requires control slightly beyond $\B$. We consider the enlarged domain
\[
\B^+=\textstyle\prod_i[\ell_i-1,0].
\]

To bound the difference between finite and fluid best responses, we require two properties of the fluid market on this enlarged domain. First, own-spend dominance must continue to hold when rivals' multipliers lie in $\B^+$, ensuring contraction of fluid best responses on the enlarged domain. Second, spending must increase sufficiently with a bidder's own multiplier near her budget, so that small spending errors imply small best-response errors. We formalize these requirements below.

\begin{assumption}[Dominant fluid reference]
\label{ass:fluid}
The fluid market $\mu$ is a probability measure, and for some $\gamma\in(0,1]$ and $\underline J>0$:
\begin{enumerate}[leftmargin=2.2em,label=(F\arabic*)]
\item \Cref{ass:smooth} holds on an open set $U\supseteq\B^+$;
\item Property \eqref{eq:uniform-thickness} holds for every bidder $i$ at every $\beta$ with $\beta_{-i}\in\B^+_{-i}$, $\beta_i\in[\ell_i,0]$ and $S_i(\beta)=B_i$;
\item $J_{ii}(\beta)\ge\underline J$ for every bidder $i$ and every $\beta\in\B^+$ with $S_i(\beta)\ge B_i/2$.
\end{enumerate}
\end{assumption}

Condition (F2) extends $(D_\gamma)$ to rival profiles in $\B^+$, while restricting bidder $i$'s own multiplier to $[\ell_i,0]$, where her fluid best response always lies. Consequently, the argument of \Cref{thm:br} still applies, and $\BR$ is a $(1-\gamma)$-contraction on $\B^+$ with fixed point $\beta^*$. Condition (F3) provides a uniform positive lower bound on the own-spend derivative near the budget. Thus, increasing bidder $i$'s log multiplier by $t$ increases her fluid spending by at least $\underline Jt$, as long as spending remains above $B_i/2$ and the profile stays in $\B^+$. This allows us to bound the difference between finite and fluid best responses when their spend maps are uniformly close.

To control spending at finite best-response iterates, we define
\[
\mathsf L= \max_{\beta\in\B^+}\max_i\sum_k|J_{ik}(\beta)|,
\qquad B_{\min}=\min_i B_i.
\] By (F1) and compactness of $\B^+$, each $S_i$ is $\mathsf L$-Lipschitz on $\B^+$ in $\|\cdot\|_\infty$. Suppose a finite instance is $\varepsilon$-close to $\mu$ on $\B^+$, and let $\beta\in\B^+$ satisfy $\|\beta-\beta^*\|_\infty\le r$. For any allocation $x$ to highest bidders, let $\hat S_i^x(\beta)$ denote bidder $i$'s total spending under $x$, normalized per item. Then
\[
|\hat S_i^x(\beta)-S_i(\beta^*)|
\le \varepsilon+\mathsf Lr.
\] Since $S_i(\beta^*)\le B_i$, normalized finite spending can exceed $B_i$ by at most $\varepsilon+\mathsf Lr$. Relative to any bidder's budget, this excess is at most $(\varepsilon+\mathsf Lr)/B_{\min}$, which bounds the common rescaling needed to restore budget feasibility.

The following theorem combines these bounds with the contraction of fluid best responses. The lower bound on the own-spend derivative in (F3) controls the difference between finite and fluid best responses, while the contraction implied by (F2) bounds the accumulation of these errors over iterations. A final common rescaling yields a budget-feasible approximate \sppe.

\begin{restatable}[Finite best-response dynamics and approximate equilibria]{theorem}{thmfinite}\label{thm:finite}
Under \Cref{ass:fluid}, suppose $I$ is $\varepsilon$-close to $\mu$ on $\B^+$, with $\varepsilon\le B_{\min}/2$ and $\varepsilon\le\underline J$, and set $\zeta=\varepsilon/\underline J\le1$. Then:
\begin{enumerate}[leftmargin=2em,label=(\roman*)]
\item \emph{Best-response error.} $\abs{\widehat{\BR}_i(\beta_{-i})-\BR_i(\beta_{-i})}\le\zeta$ for every bidder $i$ and every $\beta_{-i}\in\B^+_{-i}$.
\item \emph{Distance to the fluid equilibrium.} Finite best-response dynamics $\beta^{t+1}=\widehat{\BR}(\beta^t)$ started at $\beta^0=0$ stay in $\prod_i[\ell_i-\zeta,0]\subseteq\B^+$ and satisfy
\[
\norm{\beta^t-\beta^*}_\infty\le(1-\gamma)^t\max_i\abs{\ell_i}+\zeta/\gamma .
\]
\item \emph{A budget-feasible approximate equilibrium.} Let $t\ge t_0=\lceil\gamma^{-1}\log(\gamma\max_i\abs{\ell_i}/\zeta)\rceil$ and $r=\norm{\beta^t-\beta^*}_\infty$. Fix any allocation $x$ of each item to a highest bidder at $\beta^t$, let $\hat S^x_i(\beta^t)$ be the resulting total spending of bidder $i$, normalized by $m$, and set
\[
\Lambda=\max\Bigl\{1,\ \max_i\hat S^x_i(\beta^t)/B_i\Bigr\}.
\] Then the log multipliers $\beta^t-\log\Lambda$, with the same allocation, form a $(0,\phi)$-approximate \sppe of $I$, where
\[
\phi=\frac{2(\varepsilon+\mathsf Lr)}{B_{\min}}+r
\le\frac{2\varepsilon\chi_\mu}{\tilde v},
\qquad
\chi_\mu=
\frac{\tilde v}{B_{\min}}
\left(1+\frac{2\mathsf L}{\gamma\underline J}\right)
+\frac{\tilde v}{\gamma\underline J},
\] and $\tilde v>0$ is an arbitrary normalization constant.
\end{enumerate}
\end{restatable}

By \Cref{thm:finite}(iii), the approximate-equilibrium error satisfies $\phi\le 2\varepsilon\chi_\mu/\tilde v$, where $\varepsilon/\tilde v$ is the uniform spend-map approximation error of \Cref{def:close}, normalized by $\tilde v$. The dimensionless parameter $\chi_\mu$ captures how this normalized error is amplified when budgets or own-spend derivatives are small, or when fluid best responses contract slowly. Thus, under the conditions of \Cref{thm:finite}, a uniform spend-map error $\varepsilon\le\tilde v/(2\chi_\mu n^c)$ suffices to obtain a $(0,n^{-c})$-approximate \sppe.

For random instances with valuations bounded by $\bar v$, we take $\tilde v=\bar v$. Hence, for $m$ items drawn independently from $\mu$, \Cref{lem:vc} guarantees $\varepsilon_m$-closeness in the sense of \Cref{def:close} with probability at least $1-\varrho$. The normalized error bound $\varepsilon_m/\bar v$ decreases at rate $m^{-1/2}$ up to logarithmic and dimensional factors. Substituting this bound yields the following sample-size guarantee.

\begin{restatable}
[Inverse-polynomial approximate equilibria for random instances]
{corollary}{corckk}\label{cor:ckk} Under \Cref{ass:fluid}, suppose $0\le v_i\le\bar v$ for every bidder $i$, and fix $c>0$ and $\varrho\in(0,1]$. Let $\chi_\mu$ be defined as in \Cref{thm:finite}(iii) with $\tilde v=\bar v$. If
\[
m\ge 128\,\chi_\mu^2 n^{2c}\bigl(2n\log(m+1)+\log(16n/\varrho)\bigr),
\] then, with probability at least $1-\varrho$ over $m$ independent draws from $\mu$, every rescaled finite best-response iterate of \Cref{thm:finite}(iii) with $t\ge t_0$ is a $(0,n^{-c})$-approximate \sppe. Here $t_0$ is defined as in \Cref{thm:finite}(iii) with $\varepsilon=\varepsilon_m$, and satisfies $t_0=O(\gamma^{-1}\log m)$ for fixed fluid-market parameters. Each best-response round takes $O(nm\log m)$ time.
\end{restatable}

Although \Cref{cor:ckk} concerns randomly sampled instances, the finite best-response dynamics can be computed without knowledge of the underlying fluid market. Starting from $\beta^0=0$, each update $\beta^{t+1}=\widehat{\BR}(\beta^t)$ is determined by the bid thresholds of the realized items, while the rescaling in \Cref{thm:finite}(iii) depends only on finite-market spending. Thus, the existence of a fluid market satisfying \Cref{ass:fluid} and uniformly approximating the finite spend maps suffices to establish the approximation guarantee, without requiring that market as input. This motivates a class of finite instances for which a $(0,n^{-c})$-approximate \sppe can be computed in polynomially many best-response iterations.

Fix $c>0$ and a polynomial $\mathsf P$. Let $\mathcal I_{\mathsf P,c}$ be the class of finite instances $I$, with $n$ bidders and $m$ items, for which there exists a fluid market $\mu$ satisfying \Cref{ass:fluid}, with per-item budgets equal to those of $I$ divided by $m$, such that $I$ is $\varepsilon$-close to $\mu$ on $\B^+$, where
\[
\varepsilon=
\frac{\tilde v n^{-c}}{2\chi_\mu},
\qquad
t_0=
\left\lceil
\gamma^{-1}
\log\left(
\frac{\gamma\underline J\max_i|\ell_i|}
{\varepsilon}
\right)
\right\rceil
\le\mathsf P(n,m).
\] The constants and the domain $\B^+$ are those associated with $\mu$, and $\chi_\mu$ is defined as in \Cref{thm:finite}(iii) for any fixed normalization constant $\tilde v>0$. Since $\chi_\mu\ge\tilde v/B_{\min}$ and $\chi_\mu\ge\tilde v/(\gamma\underline J)$, this $\varepsilon$ also satisfies $\varepsilon\le B_{\min}/2$ and $\varepsilon\le\underline J$, as required by \Cref{thm:finite}.

For every $I\in\mathcal I_{\mathsf P,c}$, the bound on $\varepsilon$ ensures that the approximation error in \Cref{thm:finite}(iii) is at most $n^{-c}$, while the bound on $t_0$ ensures that this guarantee is attained within polynomially many iterations. Since these iterations and the rescaling can be computed from the finite instance, we obtain the following algorithmic and complexity consequences.

\begin{restatable}
[Polynomial-time approximation for fluid-approximable instances]
{corollary}{corhard}\label{cor:hard} There is a polynomial-time algorithm that, for every $I\in\mathcal I_{\mathsf P,c}$, computes a $(0,n^{-c})$-approximate \sppe without knowledge of the fluid market $\mu$. Consequently, unless $\mathrm{PPAD}\subseteq\mathrm P$, no reduction establishing PPAD-hardness of $(n^{-c},n^{-c})$-approximate \sppe \citep{chen2024complexity} can produce only instances in $\mathcal I_{\mathsf P,c}$.
\end{restatable}

The algorithm runs finite best-response dynamics from $\beta^0=0$. At each iterate, it selects an allocation to highest bidders, rescales the multipliers as in \Cref{thm:finite}(iii), and checks whether the resulting allocation and multipliers satisfy the conditions of a $(0,n^{-c})$-approximate \sppe. The updates, rescaling, and checks can all be performed in polynomial time. For every $I\in\mathcal I_{\mathsf P,c}$, \Cref{thm:finite}(iii) guarantees that the checks succeed by iteration $t_0\le\mathsf P(n,m)$. Hence, the algorithm need not determine membership in $\mathcal I_{\mathsf P,c}$ or construct a fluid market witnessing it.

\begin{remark}[Scope of the hardness result]
\label{rem:hardness}
As described in \Cref{rem:finite}, finite spend maps have no local displacement term between changes in winning bidders. This property also holds for instances in $\mathcal I_{\mathsf P,c}$ and therefore does not by itself imply computational hardness. Our approximation guarantee instead relies on uniform closeness to a fluid spend map satisfying \Cref{ass:fluid}: this bounds the finite best-response errors, while contraction bounds their cumulative effect on the distance to the fluid equilibrium. Thus, \Cref{cor:hard} restricts the instances that can arise in a PPAD-hardness reduction, but does not characterize the computational complexity of instances outside $\mathcal I_{\mathsf P,c}$.
\end{remark}

\section{Conclusion}

In this work, we identify conditions on the structure of competition in second-price auction markets under which the pacing equilibrium is unique and simple pacing dynamics converge to it. The central insight is that the stability of pacing depends on the extent to which competition over prices is accompanied by competition over allocations. An increase in a rival's pacing multiplier may raise a bidder's spending through higher payments but may also reduce it by displacing her from marginal items. When displacement sufficiently offsets the price effect, own-spend dominance holds. Under appropriate uniform versions of this condition, we show that best responses and projected dual descent are contractions and that per-auction stochastic pacing converges almost surely.

Our results suggest that the difficulties associated with second-price pacing can be attributed, at least in part, to market structures in which bidders influence each other's payments without effectively competing for allocations. In logit markets, we show that sufficiently many rivals with comparable bids on each item type guarantee uniqueness and convergence. Thus, what matters is not the total number of bidders, but the extent of competition for individual items. Our analysis also identifies a class of finite markets that can be approximated by fluid markets satisfying own-spend dominance and for which approximate pacing equilibria can be computed in polynomial time.

Several directions remain open. First, extending the analysis beyond the logit model and deriving sufficient conditions for own-spend dominance under alternative valuation distributions would broaden the practical applicability of our results. Second, own-spend dominance is only a sufficient condition, and it would be valuable to understand to what extent uniqueness and convergence continue to hold when it is violated. Third, since price and displacement effects can be characterized in terms of the two highest bids and the distribution of their gaps, these effects could in principle be estimated from auction data. This would allow platforms to assess whether their markets satisfy own-spend dominance and identify market segments potentially susceptible to instability.

\bibliographystyle{plainnat}
\bibliography{refs}

@article{chen2024complexity,
  title={The complexity of pacing for second-price auctions},
  author={Chen, Xi and Kroer, Christian and Kumar, Rachitesh},
  journal={Mathematics of Operations Research},
  volume={49},
  number={4},
  pages={2109--2135},
  year={2024},
  publisher={INFORMS}
}

@article{conitzer2022multiplicative,
  title={Multiplicative pacing equilibria in auction markets},
  author={Conitzer, Vincent and Kroer, Christian and Sodomka, Eric and Stier-Moses, Nicolas E},
  journal={Operations Research},
  volume={70},
  number={2},
  pages={963--989},
  year={2022},
  publisher={INFORMS}
}

@article{conitzer2022pacing,
  title={Pacing equilibrium in first price auction markets},
  author={Conitzer, Vincent and Kroer, Christian and Panigrahi, Debmalya and Schrijvers, Okke and Stier-Moses, Nicolas E and Sodomka, Eric and Wilkens, Christopher A},
  journal={Management Science},
  volume={68},
  number={12},
  pages={8515--8535},
  year={2022},
  publisher={INFORMS}
}

@article{balseiro2015repeated,
  title={Repeated auctions with budgets in ad exchanges: Approximations and design},
  author={Balseiro, Santiago R and Besbes, Omar and Weintraub, Gabriel Y},
  journal={Management Science},
  volume={61},
  number={4},
  pages={864--884},
  year={2015},
  publisher={INFORMS}
}

@article{balseiro2019learning,
  title={Learning in repeated auctions with budgets: Regret minimization and equilibrium},
  author={Balseiro, Santiago R and Gur, Yonatan},
  journal={Management Science},
  volume={65},
  number={9},
  pages={3952--3968},
  year={2019},
  publisher={INFORMS}
}

@article{ostrovsky2023reserve,
  title={Reserve prices in internet advertising auctions: A field experiment},
  author={Ostrovsky, Michael and Schwarz, Michael},
  journal={Journal of Political Economy},
  volume={131},
  number={12},
  pages={3352--3376},
  year={2023},
  publisher={The University of Chicago Press, Chicago, IL}
}

@article{edelman2007internet,
  title={Internet advertising and the generalized second-price auction: Selling billions of dollars worth of keywords},
  author={Edelman, Benjamin and Ostrovsky, Michael and Schwarz, Michael},
  journal={American economic review},
  volume={97},
  number={1},
  pages={242--259},
  year={2007},
  publisher={American Economic Association}
}

@article{varian2007position,
  title={Position auctions},
  author={Varian, Hal R},
  journal={international Journal of industrial Organization},
  volume={25},
  number={6},
  pages={1163--1178},
  year={2007},
  publisher={Elsevier}
}

@inproceedings{richardson2007predicting,
  title={Predicting clicks: estimating the click-through rate for new ads},
  author={Richardson, Matthew and Dominowska, Ewa and Ragno, Robert},
  booktitle={Proceedings of the 16th international conference on World Wide Web},
  pages={521--530},
  year={2007}
}

@inproceedings{mcmahan2013ad,
  title={Ad click prediction: a view from the trenches},
  author={McMahan, H Brendan and Holt, Gary and Sculley, David and Young, Michael and Ebner, Dietmar and Grady, Julian and Nie, Lan and Phillips, Todd and Davydov, Eugene and Golovin, Daniel and others},
  booktitle={Proceedings of the 19th ACM SIGKDD international conference on Knowledge discovery and data mining},
  pages={1222--1230},
  year={2013}
}

@inproceedings{he2014practical,
  title={Practical lessons from predicting clicks on ads at {Facebook}},
  author={He, Xinran and Pan, Junfeng and Jin, Ou and Xu, Tianbing and Liu, Bo and Xu, Tao and Shi, Yanxin and Atallah, Antoine and Herbrich, Ralf and Bowers, Stuart and others},
  booktitle={Proceedings of the eighth international workshop on data mining for online advertising},
  pages={1--9},
  year={2014}
}

@inproceedings{perlich2012bid,
  title={Bid optimizing and inventory scoring in targeted online advertising},
  author={Perlich, Claudia and Dalessandro, Brian and Hook, Rod and Stitelman, Ori and Raeder, Troy and Provost, Foster},
  booktitle={Proceedings of the 18th ACM SIGKDD international conference on Knowledge discovery and data mining},
  pages={804--812},
  year={2012}
}

@article{mcfadden1972conditional,
  title={Conditional logit analysis of qualitative choice behavior},
  author={McFadden, Daniel},
  year={1972}
}

@article{bertsekas1983distributed,
  title={Distributed asynchronous computation of fixed points},
  author={Bertsekas, Dimitri P},
  journal={Mathematical Programming},
  volume={27},
  number={1},
  pages={107--120},
  year={1983},
  publisher={Springer}
}

@book{bertsekas2015parallel,
  title={Parallel and distributed computation: numerical methods},
  author={Bertsekas, Dimitri and Tsitsiklis, John},
  year={2015},
  publisher={Athena Scientific}
}

@book{devroye1996probabilistic,
  title={A probabilistic theory of pattern recognition},
  author={Devroye, Luc and Gy{\"o}rfi, L{\'a}szl{\'o} and Lugosi, G{\'a}bor},
  volume={31},
  year={1996},
  publisher={springer, New York}
}

@book{kushner2003stochastic,
  title={Stochastic approximation and recursive algorithms and applications},
  author={Kushner, Harold J and Yin, G George},
  year={2003},
  publisher={Springer}
}

@book{nagurney2012projected,
  title={Projected dynamical systems and variational inequalities with applications},
  author={Nagurney, Anna and Zhang, Ding},
  year={2012},
  publisher={Springer Science \& Business Media}
}

@inproceedings{chen2025constant,
  title={Constant inapproximability of pacing equilibria in second-price auctions},
  author={Chen, Xi and Li, Yuhao},
  booktitle={International Conference on Web and Internet Economics},
  pages={294--306},
  year={2025},
  organization={Springer}
}

@article{babaioff2020non,
  title={Non-quasi-linear agents in quasi-linear mechanisms},
  author={Babaioff, Moshe and Cole, Richard and Hartline, Jason and Immorlica, Nicole and Lucier, Brendan},
  journal={arXiv preprint arXiv:2012.02893},
  year={2020}
}

@article{afiouni2026pacing,
  title={Pacing equilibria in abstract mechanisms},
  author={Afiouni, Salam and Kroer, Christian},
  journal={arXiv preprint arXiv:2609.27285},
  year={2026}
}

@inproceedings{borgs2007dynamics,
  title={Dynamics of bid optimization in online advertisement auctions},
  author={Borgs, Christian and Chayes, Jennifer and Immorlica, Nicole and Jain, Kamal and Etesami, Omid and Mahdian, Mohammad},
  booktitle={Proceedings of the 16th international conference on World Wide Web},
  pages={531--540},
  year={2007}
}

@article{gaitonde2022budget,
  title={Budget pacing in repeated auctions: Regret and efficiency without convergence},
  author={Gaitonde, Jason and Li, Yingkai and Light, Bar and Lucier, Brendan and Slivkins, Aleksandrs},
  journal={arXiv preprint arXiv:2205.08674},
  year={2022}
}

@misc{meta_pacing,
	title = {Meta | About {Pacing}},
	url = {https://www.facebook.com/business/help/1754368491258883},
	language = {en},
	urldate = {2026-02-08},
        year = {2026},
        organization= {Meta},
        author={{Meta}},
	journal = {Meta Business Help Center},
}

@misc{xie2021twitterpacing,
  author       = {Xie, Eddie},
  title        = {How We Built {Twitter}'s Highly Reliable Ads Pacing Service},
  year         = {2021},
  month        = dec,
  howpublished = {Twitter Engineering Blog},
  url          = {https://blog.x.com/en/engineering/how-we-built-twitter-s-highly-reliable-ads-pacing-service},
  note         = {December 16, 2021}
}

@misc{linkedinhelp_talentinsights_a422101,
  title        = {LinkedIn Help: Talent Insights Article A422101},
  author       = {{LinkedIn}},
  year         = {2026},
  howpublished = {\url{https://www.linkedin.com/help/talent-insights/answer/a422101}},
  organization         = {LinkedIn.com}
}

@misc{adobe2026pacing,
  author = {{Adobe}},
  title = {How Advertising {DSP} Optimizes Your Campaigns},
  year         = {2026},
  howpublished = {\url{https://experienceleague.adobe.com/en/docs/advertising/dsp/optimization/optimization-how-dsp-optimizes-campaigns}},
  note = {Accessed November 2026}
}

@misc{meta2026auction,
  author = {{Meta}},
  title = {How {Meta}'s Ads System Matches Your Ad
           to the Right Person at the Right Time},
  year = {2026},
  month = oct,
  howpublished = {\url{https://pt-br.facebook.com/business/news/how-metas-ads-system-matches-your-ad-to-the-right-person}}
}

@misc{linkedin2026auction,
  author = {{LinkedIn}},
  title = {Bidding Strategy Guides},
  year = {2026},
  howpublished = {\url{https://business.linkedin.com/br/pt/advertise/resources/ebooks-and-guides/bidding-guides}},
  note = {Accessed November 2026}
}

@misc{linkedin2026pacing,
  author = {{LinkedIn}},
  title = {Making the Most of Your Budget},
  year = {2026},
  howpublished = {\url{https://business.linkedin.com/advertise/ads/best-practices/maximize-your-budget}},
  note = {Accessed November 2026}
}

@misc{x2026bidding,
  author = {{\ensuremath{\mathbb{X}}}},
  title = {Bidding and Auctions {FAQs}},
  year = {2026},
  howpublished = {\url{https://help.x.com/en/business-and-advertising/bidding-and-auctions-faqs}},
  note = {Accessed November 2026}
}

@article{gao2022bidding,
  title={Bidding agent design in the {LinkedIn} ad marketplace},
  author={Gao, Yuan and Yang, Kaiyu and Chen, Yuanlong and Liu, Min and Karoui, Noureddine El},
  journal={arXiv preprint arXiv:2202.12472},
  year={2022}
}

@inproceedings{balseiro2023robust,
  title={Robust budget pacing with a single sample},
  author={Balseiro, Santiago R and Kumar, Rachitesh and Mirrokni, Vahab and Sivan, Balasubramanian and Wang, Di},
  booktitle={International Conference on Machine Learning},
  pages={1636--1659},
  year={2023},
  organization={PMLR}
}

\appendix
\section{Proofs}\label{app:proofs}

\subsection{Proofs for \texorpdfstring{\Cref{sec:model}}{Section 2}}

\lembasic*

\begin{proof}
(i) A winning bidder pays at most its own bid, so $S_i(\beta)\le\int e^{\beta_i}v_i\dd\mu=e^{\beta_i}V_i$. If an equilibrium had $\beta_i<\ell_i$, then $S_i(\beta)<e^{\ell_i}V_i\le B_i$: the bidder would be paced and underspend, contradicting condition~(d).

(ii) Hold rival bids fixed and raise $\beta_i$. Bidder $i$ keeps every item it previously won, at the same price, and may win additional items.

(iii) Adding $t$ to every log multiplier multiplies every bid by $e^t$. Rankings do not change and every payment scales by $e^t$, so $S(\beta+t\one)=e^tS(\beta)$.

The same arguments apply to first-price payments. In (i) and (iii) they apply verbatim, since first-price payments equal the winner's bid and scale by $e^t$. In (ii), raising $\beta_i$ also raises the payment on items bidder $i$ already wins, which only strengthens the conclusion.
\end{proof}

\lemfixed*

\begin{proof}
By \Cref{subsec:spend-map}, $\beta\in\B$ is an \sppe if and only if $S_i(\beta)\le B_i$ for all $i$, with equality whenever $\beta_i<0$. (An allocation satisfying (a) and (b) exists: give each item to a highest bidder, breaking ties by index.)

Suppose $\beta=\BR(\beta)$. If $\beta_i=0$, feasibility of the best response gives $S_i(\beta)\le B_i$. If $\beta_i<0$, the response must exhaust the budget, since otherwise continuity would allow a slightly larger feasible multiplier. So $\beta$ is an \sppe.

Conversely, let $\beta$ be an \sppe. If $\beta_i=0$, it is the largest feasible log multiplier, so $\BR_i(\beta_{-i})=0$. If $\beta_i<0$, then $S_i(\beta)=B_i$. Let $x^*=\BR_i(\beta_{-i})$. Since $\beta_i$ is feasible, $x^*\ge\beta_i$. If $x^*>\beta_i$, then by monotonicity $B_i=S_i(\beta)\le S_i(x,\beta_{-i})\le S_i(x^*,\beta_{-i})\le B_i$ for all $x\in[\beta_i,x^*]$, so spend is constant at $B_i$ on this interval. This contradicts $J_{ii}(\beta)>0$, which holds because $S_i(\beta)=B_i$. Hence $\beta_i=\BR_i(\beta_{-i})$.
\end{proof}

\subsection{Proofs for \texorpdfstring{\Cref{sec:thickness}}{Section 3}}

\lemfpdominance*

\begin{proof}
Holding bidder $i$'s multiplier fixed, an increase in a rival's multiplier cannot change the payments on items that $i$ continues to win, but may cause $i$ to lose some items. Hence $J_{ik}^{\mathrm{FP}}\le 0$ for all $k\ne i$. The scaling identity holds under first price: $J_{ii}^{\mathrm{FP}} +\sum_{k\ne i}J_{ik}^{\mathrm{FP}} =S_i^{\mathrm{FP}}.$ Since all cross derivatives are nonpositive, it follows that $J_{ii}^{\mathrm{FP}}-\sum_{k\ne i}|J_{ik}^{\mathrm{FP}}| =J_{ii}^{\mathrm{FP}} +\sum_{k\ne i}J_{ik}^{\mathrm{FP}} =S_i^{\mathrm{FP}}.$
\end{proof}

\lemmargin*

\begin{proof}
If $(D_\gamma)$ holds for some $\gamma>0$, then \eqref{eq:uniform-thickness} immediately implies strict diagonal dominance in row $i$ at every $\beta\in\mathcal B$ with $S_i(\beta)=B_i$.

Conversely, suppose pointwise own-spend dominance holds for every bidder $i$ at every $\beta\in\mathcal B$ with $S_i(\beta)=B_i$, and fix $i$. At such profiles, strict diagonal dominance implies $J_{ii}>0$, so
\[
    g_i(\beta)
    :=\frac{
        J_{ii}(\beta)-\sum_{k\ne i}|J_{ik}(\beta)|
    }{J_{ii}(\beta)}
\] is well defined and positive. It is continuous because the spend Jacobian is continuous. The set of $\beta\in\mathcal B$ with $S_i(\beta)=B_i$ is a closed subset of the compact box $\mathcal B$, hence compact. If this set is nonempty, $g_i$ attains a minimum $\gamma_i\in(0,1]$ on it; if it is empty, put $\gamma_i=1$. Then $(D_\gamma)$ holds with $\gamma=\min_i\gamma_i>0$.

The argument for $(D_\gamma^+)$ is identical, with the set of budget-exhausting profiles replaced by $\mathcal B$.
\end{proof}

\subsection{Proofs for \texorpdfstring{\Cref{sec:jacobian}}{Section 4}}

\lemjacobian*

\begin{proof}
Fix $\beta\in U$. Write $y_j=\log b_j$ and $\bar y_i=\max_{j\ne i}y_j$.

\emph{Step 0: null ties and continuity.} A positive-price tie for first place between exactly two bidders $i,k$ would be an atom at $0$ of $\nu^\beta_{ik}$, which (R1) excludes. A tie among three or more bidders has $y_{(2)}=y_{(3)}$, which the second part of (R2) excludes. So positive-price ties for first place are $\mu$-null, spend does not depend on tie breaking, and bidder $i$ wins exactly when $y_i>\bar y_i$, paying $e^{\bar y_i}$:
\[
S_i(\beta)=\int\ind\{y_i>\bar y_i\}e^{\bar y_i}\dd\mu.
\] For continuity, let $\beta^n\to\beta$ in $U$. On every item with no positive-price tie for first place at $\beta$, the winner at $\beta^n$ eventually equals the winner at $\beta$, and the price converges. Every payment is at most $e^{\max_j\beta^n_j}\max_jv_j$, which is bounded by a constant multiple of $\max_j v_j\le\sum_j v_j$ and so is integrable because $V_j<\infty$ for every $j$. Dominated convergence shows that $S$ is continuous on $U$. The same argument applies to $S^{FP}$.

\emph{Step 1: raise a rival's multiplier.} Fix $k\ne i$ and increase $\beta_k$ by $t>0$. The highest rival log-bid becomes
\[
\bar y_i'=\max\{\bar y_i,y_k+t\}\in[\bar y_i,\bar y_i+t].
\] Only items that $i$ initially wins can affect the change in its spending. Split these items into four cases:
\begin{enumerate}[leftmargin=2em,label=(\alph*)]
\item \emph{$k$ is second and $i$ keeps the item.} If $y_i-y_k>t$, the price rises from $e^{y_k}$ to $e^{y_k+t}$. The change is $(e^t-1)e^{y_k}$.
\item \emph{$k$ is second and overtakes $i$.} If $0<y_i-y_k\le t$, bidder $i$ stops paying. The total lost spend is
\[
\nu_{ik}^{\beta}((0,t])=tD_{ik}+o(t)
\] by (R1).
\item \emph{$k$ starts below second place and becomes the price-setter, while $i$ keeps the item.} The price increase is at most $e^{\bar y_i}(e^t-1)=(e^t-1)p$. Such items have $k$ within $t$ of the old runner-up, so they lie in $\{0\le y_{(2)}-y_{(3)}<t\}$. As $t\downarrow0$ these sets decrease to $\{y_{(2)}=y_{(3)}\}$, which has $p$-weighted measure zero by (R2). Dominated convergence gives $\int p\,\ind\{y_{(2)}-y_{(3)}<t\}\dd\mu=o(1)$, so the total contribution is $(e^t-1)\,o(1)=o(t)$.
\item \emph{$k$ starts below second place and overtakes $i$.} All three leading log-bids must lie within $t$ of one another. The first part of (R2) makes the lost spend $o(t)$.
\end{enumerate}
In case (a), excluding the items in case (b) changes the price-weighted mass by $O(t)$, and hence changes the price-increase term by only $O(t^2)$. Summing the four cases gives
\[
S_i(\beta+te_k)-S_i(\beta)=tP_{ik}-tD_{ik}+o(t).
\]

Now lower $\beta_k$ by $u>0$. On items where $i$ is first and $k$ second, the new price is $\max\{e^{y_k-u},e^{y_{(3)}}\}$. It equals $e^{-u}e^{y_k}$ except on $\{y_{(2)}-y_{(3)}<u\}$, whose $p$-weighted mass is $o(1)$ as in case (c). The payment change on these items is therefore $-uP_{ik}+o(u)$. Bidder $i$ also gains the items on which $k$ was first and $i$ second with $y_k-y_i<u$. The new runner-up bid lies between $e^{y_i-u}$ and $e^{y_i}$, so the new price is $e^{y_i}(1+O(u))$, and these items contribute $\nu^\beta_{ki}((0,u))(1+O(u))=uD_{ki}+o(u)$. No other item changes: $i$ cannot gain an item on which a bidder other than $k$ outbids it, and prices on items where $k$ is below second place do not move. Since $D_{ki}=D_{ik}$, both one-sided derivatives agree:
\[
J_{ik}=P_{ik}-D_{ik}.
\]

\emph{Step 2: raise the bidder's own multiplier.} Raising $\beta_i$ by $t$ leaves payments on already-won items unchanged. It adds items on which $i$ was second within $t$ of the winner $k$. At these near ties, the winning bid differs from $i$'s bid by a factor $1+O(t)$, so their total payment is
\[
t\sum_{k\ne i}D_{ki}+o(t)=t\sum_{k\ne i}D_{ik}+o(t).
\] An item on which $i$ starts below second place can be gained only if $y_{(1)}-y_{(3)}<t$. Its new price is at most $e^{y_{(1)}}\le e^tp$, so by (R2) these items contribute $o(t)$. Lowering $\beta_i$ by $u>0$ leaves the price of every item that $i$ keeps unchanged, gains nothing, and loses exactly the items on which $i$ was first within $u$ of its (a.e.\ unique) runner-up $k$. The lost payment is $\sum_{k\ne i}\nu^\beta_{ik}((0,u])=u\sum_{k\ne i}D_{ik}+o(u)$. Consequently $J_{ii}=\sum_{k\ne i}D_{ik}$. Together with Step 1, this is $J=P+L$.

\emph{Step 3: first price.} The allocation changes are the same under first price. At a tie the two bids agree, so these changes carry the same weights $D_{ik}$. The weights also agree in case (d) of Step 1, where the first-price loss is the winner's own bid $e^{y_{(1)}}$ rather than $p$: on $\{y_{(1)}-y_{(3)}\le t\}$ we have $e^{y_{(1)}}\le e^tp$, so this loss is still $o(t)$ by (R2). The payment changes differ: raising $\beta_i$ multiplies all of $i$'s existing payments by $e^t$, adding $S_i^{FP}$ to its own derivative, while raising a rival's bid has no direct effect on the prices $i$ pays. Hence
\[
J_{ii}^{FP}=S_i^{FP}+\sum_{k\ne i}D_{ik},\qquad
J_{ik}^{FP}=-D_{ik}\quad(k\ne i),
\] which gives $J^{FP}=\diag(S^{FP})+L$.

\emph{Step 4: continuous differentiability.} Dominated convergence, as in Step 0, gives continuity of $P_{ik}$ and of $S^{FP}$; (R1) gives continuity of $D_{ik}$. Thus all partial derivatives of $S$ and $S^{FP}$ exist and are continuous on $U$, so both maps are continuously differentiable there. Together with Step 0, this is \Cref{ass:smooth}.
\end{proof}

\subsection{Proofs for \texorpdfstring{\Cref{sec:contraction}}{Section 5}}

\thmbr*

\begin{proof}
We first bound the derivative of a paced best response. We then show that this bound extends across the profiles at which the bidder stops pacing.

\emph{Step 1: differentiate a paced best response.} Fix $i$ and define the relatively open set
\[
O_i=\{\beta_{-i}\in\B_{-i}:S_i(0,\beta_{-i})>B_i\}.
\] These are exactly the rival profiles at which bidder $i$ must pace. On $O_i$, its best response is a solution of
\[
S_i(\BR_i(\beta_{-i}),\beta_{-i})=B_i,
\] and $(D_\gamma)$ gives $J_{ii}>0$ at every such solution. The solution lies in $(\ell_i,0)$. It is below $0$ because $S_i(0,\beta_{-i})>B_i$. It is above $\ell_i$ because, when $\ell_i<0$, $S_i(\ell_i,\beta_{-i})<e^{\ell_i}V_i=B_i$: positive-price ties for the highest bid are $\mu$-null by \Cref{ass:no-ties}, so on almost every item with $v_i>0$, bidder $i$ either loses or pays strictly less than its bid, and $\{v_i>0\}$ has positive measure because $V_i>0$. (If $\ell_i=0$, then $O_i=\emptyset$.) The solution is unique, as in the proof of \Cref{lem:fixed}.

The map $S$ is $C^1$ on the open set $U\supset\B$, and $J_{ii}>0$ at the solution. The implicit function theorem therefore gives a $C^1$ function $x(\cdot)$ on a neighborhood $N\subseteq\R^{n-1}$ of $\beta_{-i}$ with $x(\beta_{-i})=\BR_i(\beta_{-i})$ and $S_i(x(\beta'_{-i}),\beta'_{-i})=B_i$ for $\beta'_{-i}\in N$. After shrinking $N$, $x$ takes values in $(\ell_i,0)$ and $N\cap\B_{-i}\subseteq O_i$. Uniqueness of the budget-exhausting solution then gives $\BR_i=x$ on $N\cap\B_{-i}$. Hence $\BR_i$ is $C^1$ on $O_i$ and satisfies \eqref{eq:dbr}. Since the budget-exhausting profile lies in $\B$, $(D_\gamma)$ gives
\[
\sum_{k\ne i}\Bigl|\frac{\partial\BR_i}{\partial\beta_k}\Bigr|
=\frac{\sum_{k\ne i}|J_{ik}|}{J_{ii}}
\le1-\gamma
\qquad\text{on }O_i.
\]

\emph{Step 2: continuity where pacing stops.} Outside $O_i$, the best response is zero. Let $\beta_{-i}\notin O_i$ be a limit of rival profiles $\beta^n_{-i}\in\B_{-i}$, and let $x$ be any limit point of $\BR_i(\beta^n_{-i})$. If infinitely many $\beta^n_{-i}$ lie outside $O_i$, those responses equal zero, so $x=0$ along that subsequence. Otherwise, continuity of spend gives $S_i(x,\beta_{-i})=B_i$ along a subsequence in $O_i$, while $S_i(0,\beta_{-i})\le B_i$ and $x\le0$. Since $S_i(\cdot,\beta_{-i})$ is nondecreasing, it equals $B_i$ on $[x,0]$; if $x<0$, this contradicts $J_{ii}>0$ at budget-exhausting profiles. Thus $x=0$, and the paced and unpaced branches join continuously. In particular, $\BR_i$ is continuous on $\B_{-i}$.

\emph{Step 3: a global Lipschitz bound.} Join two rival profiles $u,v\in\B_{-i}$ by a line segment, and set
\[
\Delta=\norm{v-u}_\infty,\qquad
\varphi(s)=\BR_i(u+s(v-u)),\quad 0\le s\le1.
\] The set $\{s:\varphi(s)<0\}$ is relatively open in $[0,1]$, so it is a union of relatively open intervals. On each such interval, Step 1 gives
\[
|\varphi'(s)|\le\sum_{k\ne i}\Bigl|\frac{\partial\BR_i}{\partial\beta_k}\Bigr|\,\Delta
\le(1-\gamma)\Delta,
\] and by Step 2, $\varphi=0$ at every endpoint of such an interval that lies in $(0,1)$.

Let $0\le s<s'\le1$. If $s$ and $s'$ lie in the same paced interval, integrate the derivative bound. Otherwise, bound $|\varphi(s)|$ by integrating from $s$ to the right endpoint of its interval, and bound $|\varphi(s')|$ by integrating from the left endpoint of its interval to $s'$; at an unpaced point the value is already zero. Both values are nonpositive, so their difference is at most the larger of their absolute values. In either case,
\[
|\varphi(s')-\varphi(s)|\le(1-\gamma)\Delta(s'-s).
\] Taking $s=0$ and $s'=1$ gives $|\BR_i(v)-\BR_i(u)|\le(1-\gamma)\norm{v-u}_\infty$. Taking the maximum over $i$ shows that $\BR$ is a $(1-\gamma)$-contraction of $\B$ in $\norm{\cdot}_\infty$.

\emph{Step 4: identify the limit.} The box $\B$ is complete, so Banach's fixed-point theorem gives a unique fixed point and the stated geometric bound. By \Cref{lem:fixed}, whose hypothesis is implied by $(D_\gamma)$, fixed points of $\BR$ are exactly the \sppe; by \Cref{lem:basic}, every \sppe lies in $\B$. The asynchronous conclusion follows from the convergence theorem for maximum-norm contractions \citep{bertsekas1983distributed,bertsekas2015parallel}.
\end{proof}

\corfpbr*

\begin{proof}
By \Cref{lem:first-price-diagonal-dominance},
\[
J_{ik}^{FP}\le 0,
\qquad
J_{ii}^{FP}-\sum_{k\ne i}|J_{ik}^{FP}|=S_i^{FP}.
\] At a budget-exhausting profile, $S_i^{FP}=B_i>0$, so $J^{FP}_{ii}\ge B_i>0$ and
\[
\sum_{k\ne i}|J^{FP}_{ik}|=J^{FP}_{ii}-B_i=\Bigl(1-\frac{B_i}{J^{FP}_{ii}}\Bigr)J^{FP}_{ii}\le(1-\gamma^{FP})\,J^{FP}_{ii} .
\] Bidder $i$'s budget surface $\{\beta\in\B:S^{FP}_i(\beta)=B_i\}$ is compact, and $J^{FP}_{ii}$ is continuous and at least $B_i$ on it, so $B_i/J^{FP}_{ii}$ is bounded below by a positive constant there. Hence $\gamma^{FP}\in(0,1]$, and $(D_\gamma)$ holds for $S^{FP}$ with $\gamma=\gamma^{FP}$.

At a paced best response, $\BR^{FP}_i(\beta_{-i})<0$, the first-price analogue of \eqref{eq:dbr} gives
\[
\frac{\partial\BR_i^{FP}}{\partial\beta_k}
=-\frac{J_{ik}^{FP}}{J_{ii}^{FP}}\ge 0,
\qquad
\sum_{k\ne i}
\left|
\frac{\partial\BR_i^{FP}}{\partial\beta_k}
\right|
=1-\frac{B_i}{J_{ii}^{FP}}
\le 1-\gamma^{FP}.
\] The sensitivity bound extends to boundary best responses as in the proof of \Cref{thm:br}, which uses only \Cref{ass:smooth}, \Cref{lem:basic} (which holds for $S^{FP}$) and $(D_\gamma)$, with one exception. Step~1 of that proof places a paced best response strictly above $\ell_i$ using $S_i(\ell_i,\beta_{-i})<B_i$, which under first price can hold with equality. The step goes through without it: by \Cref{lem:basic}(i), $S^{FP}_i(\ell_i,\beta'_{-i})\le B_i$ for every $\beta'_{-i}$, so if the implicit solution satisfied $x(\beta'_{-i})<\ell_i$, monotonicity would force $S^{FP}_i(\cdot,\beta'_{-i})=B_i$ on $[x(\beta'_{-i}),\ell_i]$, contradicting $J^{FP}_{ii}>0$ on the budget surface. Hence the implicit solution coincides with the best response, and $\BR^{FP}$ is a $(1-\gamma^{FP})$-contraction of $\B$. The remaining claims follow as in \Cref{thm:br}, using \Cref{lem:fixed} for $S^{FP}$.

Finally, $S^{FP}_i$ is nonincreasing in each $\beta_k$, $k\ne i$, because $J^{FP}_{ik}\le0$. Raising rival multipliers therefore enlarges the set $\{x\in[\ell_i,0]:S^{FP}_i(x,\beta_{-i})\le B_i\}$, so $\BR^{FP}_i$ is nondecreasing in $\beta_{-i}$, and $\BR^{FP}$ is isotone.
\end{proof}

\thmdd*

\begin{proof}
By $(D_\gamma^+)$, $J_{ii}$ is strictly positive throughout $\B$. Continuity and compactness therefore imply $0<J_{\min}\le J_{\max}<\infty$. In particular, admissible step sizes exist, and the modulus lies in $[0,1)$ because $\gamma J_{\min}\min_i\eta_i\le J_{\min}/J_{\max}\le1$.

We first bound the sensitivity of the update before projection. Let
\[
F(\beta)=\beta-\eta\circ(S(\beta)-B).
\] Its Jacobian is $I-\diag(\eta)J$. Since $\eta_iJ_{ii}\le1$, the diagonal entry in row $i$ is nonnegative, and by $(D^+_\gamma)$ the absolute row sum is
\[
\sum_k|\partial_kF_i|
=1-\eta_iJ_{ii}+\eta_i\sum_{k\ne i}|J_{ik}|
\le1-\eta_i\gamma J_{ii}
\le1-\gamma J_{\min}\min_k\eta_k.
\] The maximum absolute row sum is the operator norm induced by $\norm{\cdot}_\infty$. Integrating the Jacobian along the segment between two points of the convex box $\B$ therefore gives the stated Lipschitz bound for $F$. Coordinatewise clipping cannot increase maximum-norm distances, so the same bound holds for $T=\Pi_\B\circ F$.

It remains to identify the fixed point. At a fixed point of $T$, consider a coordinate with $\ell_i<0$:
\begin{itemize}[leftmargin=1.5em]
\item If $\ell_i<\beta_i<0$, projection is inactive, so $S_i=B_i$.
\item If $\beta_i=0$, the unprojected update must point upward or stay fixed, so $S_i\le B_i$.
\item If $\beta_i=\ell_i$, the unprojected update must point downward or stay fixed, so $S_i\ge B_i$. But \Cref{lem:basic}(i) also gives $S_i\le e^{\ell_i}V_i=B_i$, hence equality.
\end{itemize}
If $\ell_i=0$, that coordinate is fixed at zero and $S_i\le V_i\le B_i$ by \Cref{lem:basic}(i). Thus every fixed point satisfies budget feasibility and no unnecessary pacing, so it is an \sppe. Conversely, these conditions make every \sppe in $\B$ a fixed point, and every \sppe lies in $\B$. The contraction mapping theorem now gives uniqueness and convergence.
\end{proof}

\thmsa*

\begin{proof}
Write $g(\beta)=-c\circ(S(\beta)-B)$, let $J_{\min}=\min_{\beta\in\B}\min_iJ_{ii}(\beta)>0$ as in \Cref{thm:dd}, and set $\kappa=\gamma J_{\min}\min_ic_i$. The fixed-point argument of \Cref{thm:dd} does not use the step sizes, so its fixed-point characterization applies to the unique \sppe $\beta^*$, which exists by \Cref{thm:dd}.

\emph{Step 1: noise and the limiting ODE.} Let $\mathcal F_t$ be the $\sigma$-field generated by the items before auction $t$. Then $\beta^t$ is $\mathcal F_t$-measurable, and $\theta_t$ is independent of $\mathcal F_t$. Positive-price ties are $\mu$-null at every $\beta\in\B$ by \Cref{ass:no-ties}, and every price satisfies $p(\theta;\beta)\le p(\theta;0)$ because $\beta\le0$ lowers every bid. Hence, for any tie-breaking rule,
\[
\E[z^t\mid\mathcal F_t]=S(\beta^t),\qquad
\E[(z^t_i)^2\mid\mathcal F_t]\le\int p(\theta;0)^2\dd\mu(\theta)<\infty .
\] Thus $\beta^{t+1}=\Pi_\B[\beta^t+\epsilon_t(g(\beta^t)+M^t)]$, where $M^t$ is a martingale difference with uniformly bounded second moments. The mean field $g$ is continuous, there is no bias term, the constraint set is a hyperrectangle, and $\sum_t\epsilon_t=\infty$, $\sum_t\epsilon_t^2<\infty$. These are the hypotheses of \citet[Thm.~5.2.1]{kushner2003stochastic}. Almost surely, therefore, $\beta^t$ converges to the limit set in $\B$ of the projected ODE $\dot\beta=g(\beta)+\zeta$, where $\zeta(t)$ is the minimal force keeping $\beta(t)$ in $\B$. On a box, this ODE decouples by coordinate:
\[
\dot\beta_i=
\begin{cases}
g_i(\beta), & \ell_i<\beta_i<0,\\
\min\{g_i(\beta),0\}, & \beta_i=0>\ell_i,\\
\max\{g_i(\beta),0\}, & \beta_i=\ell_i<0,\\
0, & \ell_i=0 .
\end{cases}
\] Since $g$ is $C^1$ on a neighborhood of the compact box, it is Lipschitz on $\B$. Hence every initial point in $\B$ has a unique absolutely continuous solution $\phi_t(\beta)$, which satisfies the display for almost every $t$ \citep{nagurney2012projected}.

\emph{Step 2: the projected flow contracts in the maximum norm.} Let $\beta(\cdot)$ and $\beta'(\cdot)$ be two solutions, and write $e=\beta-\beta'$ and $V(t)=\norm{e(t)}_\infty$. Then $V$ is absolutely continuous. Fix a time $t$ at which $\beta$, $\beta'$ and $V$ are differentiable and the coordinatewise equations hold; almost every $t$ has this property. Suppose $V(t)>0$ and choose $i$ with $\abs{e_i(t)}=V(t)$. By symmetry we may assume $e_i(t)>0$. Then $\beta_i(t)>\ell_i$ and $\beta'_i(t)<0$, so the display gives
\[
\dot\beta_i\le g_i(\beta),\qquad \dot\beta'_i\ge g_i(\beta').
\] Because $e_i\le V$ with equality at $t$, we have $V'(t)=\dot e_i(t)\le g_i(\beta)-g_i(\beta')$. Integrating along the segment $\beta^s=\beta'+se\in\B$ gives
\[
g_i(\beta)-g_i(\beta')
=-c_i\int_0^1\Bigl[J_{ii}(\beta^s)e_i+\sum_{k\ne i}J_{ik}(\beta^s)\,e_k\Bigr]\dd s .
\] Since $\abs{e_k}\le V=e_i$, $(D^+_\gamma)$ shows that, at every $\beta^s$, the bracket is at least
\[
\Bigl(J_{ii}-\sum_{k\ne i}\abs{J_{ik}}\Bigr)V\ge\gamma J_{ii}V\ge\gamma J_{\min}V .
\] Hence $V'(t)\le-c_i\gamma J_{\min}V(t)\le-\kappa V(t)$. If $V(t)=0$, then $t$ minimizes $V$, so $V'(t)=0$. Thus $(e^{\kappa t}V(t))'\le0$ almost everywhere, and absolute continuity gives
\[
\norm{\phi_t(\beta)-\phi_t(\beta')}_\infty\le e^{-\kappa t}\norm{\beta-\beta'}_\infty .
\]

\emph{Step 3: identify the limit set.} The \sppe $\beta^*$ is a rest point of the ODE. By the fixed-point characterization in the proof of \Cref{thm:dd}, $g_i(\beta^*)=0$ when $\ell_i<\beta^*_i<0$ or $\beta^*_i=\ell_i<0$, and $g_i(\beta^*)\ge0$ when $\beta^*_i=0$. In every case the display gives $\dot\beta_i=0$. Taking $\beta'\equiv\beta^*$ in Step 2 shows that every trajectory in $\B$ converges to $\beta^*$ exponentially, uniformly in its starting point. The limit set of the ODE in $\B$ is therefore $\{\beta^*\}$, and $\beta^t\to\beta^*$ almost surely.
\end{proof}

\subsection{Proofs for \texorpdfstring{\Cref{sec:logit}}{Section 6}}

\lemlogit*

\begin{proof}
Each condition of \Cref{ass:regularity} and each formula is preserved by finite sums over types, so it suffices to consider one type of unit mass. We suppress its index. Let $Z_k=\beta_k+a_k+\sigma\varepsilon_k$ denote bidder $k$'s log-bid. Its distribution function and density are
\[
F_k(z)=e^{-q_ke^{-z/\sigma}},\qquad
f_k(z)=\frac{q_k}{\sigma}e^{-z/\sigma}e^{-q_ke^{-z/\sigma}}.
\] By independence, the maximum of the log-bids in a set $A$ of bidders has distribution function $e^{-q_Ae^{-z/\sigma}}$, where $q_A=\sum_{k\in A}q_k$. All the calculations below use the change of variables $s=e^{-z/\sigma}$, under which $e^z=s^{-\sigma}$, together with the gamma integral
\[
\int_0^\infty s^{c-1}e^{-bs}\dd s=\Gamma(c)\,b^{-c}\qquad(b,c>0).
\] The restriction $\sigma<1$ ensures that the integrals with $c=1-\sigma$ are finite.

\emph{Step 1: regularity.} Total values are finite: $V_k=\E[e^{a_k+\sigma\varepsilon_k}]=\Gamma(1-\sigma)e^{a_k}<\infty$ for every $k$.

(R1): For $g\ge0$, the event that $i$ and $k$ are the two highest bidders with $Z_k=z$ and $Z_i=z+g$ has price $e^z$. Its density is $f_k(z)f_i(z+g)\prod_{l\ne i,k}F_l(z)$, so
\[
H_{ik}(g;\beta)=\int e^zf_k(z)f_i(z+g)\prod_{l\ne i,k}F_l(z)\dd z
=\frac{\Gamma(2-\sigma)}{\sigma}\,q_iq_k\,e^{-g/\sigma}\bigl(q_{-i}+q_ie^{-g/\sigma}\bigr)^{\sigma-2}.
\] For $g<0$, $H_{ik}(g;\beta)=H_{ki}(-g;\beta)$. Both expressions are continuous in $(g,\beta)$ and agree at $g=0$.

(R2): Each density satisfies $f_l(z)=\sigma^{-1}xe^{-x}\le(e\sigma)^{-1}$, where $x=q_le^{-z/\sigma}$. On $\{y_{(1)}-y_{(3)}\le t\}$, some bidder $u$ has the top log-bid $z$, two others have log-bids in $[z-t,z]$, and $p\le e^z$. Therefore
\[
\int p\,\ind\{y_{(1)}-y_{(3)}\le t\}\dd\mu
\le\sum_{u}\binom{n-1}{2}\Bigl(\frac{t}{e\sigma}\Bigr)^2\int e^zf_u(z)\dd z
=\binom{n-1}{2}\frac{t^2}{(e\sigma)^2}\,\Gamma(1-\sigma)\sum_uq_u^{\sigma},
\] which is $O(t^2)$. Finally, the log-bids are independent with densities, so ties occur with probability zero. Hence \Cref{ass:regularity} holds at every $\beta\in\R^n$.

\emph{Step 2: spend.} Let $Y=\max_{k\ne i}Z_k$ be the highest rival log-bid. Conditional on $Y$, bidder $i$ pays $e^Y$ with probability $1-F_i(Y)$. The variable $e^{-Y/\sigma}$ has density $q_{-i}e^{-q_{-i}s}$. Therefore
\[
S_i=\E\bigl[e^Y(1-F_i(Y))\bigr]
=q_{-i}\int_0^\infty s^{-\sigma}\bigl(e^{-q_{-i}s}-e^{-\bar qs}\bigr)\dd s
=\Gamma(1-\sigma)\bigl[q_{-i}^{\sigma}-q_{-i}\bar q^{\,\sigma-1}\bigr].
\]

\emph{Step 3: price effect.} For $k$ to set $i$'s price at log-bid $z$, bidder $i$ must bid above $z$ and every bidder other than $i$ and $k$ must bid below $z$. Weighting this event by its price gives
\[
P_{ik}
=\int e^z f_k(z)\prod_{l\ne i,k}F_l(z)\,[1-F_i(z)]\dd z
=q_k\int_0^\infty s^{-\sigma}\bigl(e^{-q_{-i}s}-e^{-\bar qs}\bigr)\dd s
=\Gamma(1-\sigma)\,q_k\bigl[q_{-i}^{\sigma-1}-\bar q^{\,\sigma-1}\bigr].
\]

\emph{Step 4: displacement effect.} By Step 1 and \eqref{eq:D},
\[
D_{ik}=H_{ik}(0;\beta)
=\frac{q_iq_k}{\sigma}\int_0^\infty s^{1-\sigma}e^{-\bar qs}\dd s
=\frac{\Gamma(2-\sigma)}{\sigma}\,q_iq_k\bar q^{\,\sigma-2}.
\]

\emph{Step 5: the ratio.} Dividing the formulas of Steps 3 and 4 cancels $q_k$. Using $\Gamma(2-\sigma)=(1-\sigma)\Gamma(1-\sigma)$, $q_i=w_i\bar q$ and $q_{-i}=(1-w_i)\bar q$,
\[
\frac{P_{ik}}{D_{ik}}
=\frac{\sigma}{1-\sigma}\cdot\frac{q_{-i}^{\sigma-1}-\bar q^{\,\sigma-1}}{q_i\bar q^{\,\sigma-2}}
=\frac{\sigma}{1-\sigma}\cdot\frac{(1-w_i)^{-(1-\sigma)}-1}{w_i}
=\rho(\sigma,w_i).
\qedhere
\]
\end{proof}

\lemrho*

\begin{proof}
We use the integral representation
\[
\rho(\sigma,w)=\sigma\int_0^1(1-tw)^{-(2-\sigma)}\dd t ,\qquad w\in[0,1),
\] which holds for $w\in(0,1)$ because
\[
\sigma\int_0^1(1-tw)^{-(2-\sigma)}\dd t
=\frac{\sigma}{(1-\sigma)w}\Bigl[(1-tw)^{-(1-\sigma)}\Bigr]_{t=0}^{t=1}
=\frac{\sigma}{1-\sigma}\cdot\frac{(1-w)^{-(1-\sigma)}-1}{w},
\] and which defines the continuous extension at $w=0$. For each $t\in(0,1]$, the integrand is continuous and strictly increasing in $w\in[0,1)$, and it is bounded on $[0,\bar w]\times[0,1]$ for every $\bar w<1$. Hence $\rho(\sigma,\cdot)$ is continuous and strictly increasing on $[0,1)$, with value $\sigma$ at $w=0$. As $w\uparrow1$, monotone convergence gives $\rho(\sigma,w)\to\sigma\int_0^1(1-t)^{-(2-\sigma)}\dd t=\infty$, since $2-\sigma>1$. Because $\sigma<1<2$, the intermediate value theorem gives unique solutions $w_1(\sigma)<w_2(\sigma)$ of $\rho=1$ and $\rho=2$.
\end{proof}

\propdominance*

\begin{proof}
Write $\rho_{ij}=\rho(\sigma,w_{ij})$. By \Cref{lem:logit}, $J_{ik}=\sum_j\mu_j\bigl(P^{(j)}_{ik}-D^{(j)}_{ik}\bigr)$, and on each type with $w_{ij}>0$, $D^{(j)}_{ik}=P^{(j)}_{ik}/\rho_{ij}$ because $\rho_{ij}\ge\sigma>0$; on a type with $w_{ij}=0$, both effects vanish. Hence $P^{(j)}_{ik}-D^{(j)}_{ik}=(1-1/\rho_{ij})P^{(j)}_{ik}$. Since the positive part of a sum is at most the sum of the positive parts,
\[
[J_{ik}]_+\le\sum_j\mu_j\bigl[(1-1/\rho_{ij})P^{(j)}_{ik}\bigr]_+=\sum_j\mu_je_{ij}P^{(j)}_{ik}.
\] On each type, the rivals partition bidder $i$'s spend according to who sets its price, so $\sum_{k\ne i}P^{(j)}_{ik}=S^{(j)}_i$. Summing over $k$ gives $C_i\le\sum_j\mu_je_{ij}S^{(j)}_i=S_i\sum_js_{ij}e_{ij}$. The margin follows from \eqref{eq:slack}.
\end{proof}

\corlogithalf*

\begin{proof}
Both parts bound $\sum_js_{ij}e_{ij}$ and apply \Cref{prop:dominance}, using $\sum_js_{ij}=1$ and $0\le e_{ij}\le1$. By \Cref{lem:rho}, $e_{ij}$ is nondecreasing in $w_{ij}$, vanishes for $w_{ij}\le w_1(\sigma)$, and is below $1/2$ for $w_{ij}<w_2(\sigma)$.

(a) Every type with $s_{ij}>0$ has $e_{ij}<1/2$, so $\sum_js_{ij}e_{ij}<1/2$.

(b) If $s>1/2$ is the spending share of types with $w_{ij}\le w_1(\sigma)$, then $\sum_js_{ij}e_{ij}\le1-s<1/2$.

\end{proof}

\thmlogit*

\begin{proof}
Each $w_{ij}$ is continuous and strictly less than $1$. Since $\B$ is compact, \(w_{ij}(\beta)\) is continuous in \(\beta\), and there are finitely many bidders and item types, then $\omega<1$. By \Cref{lem:logit}, \Cref{ass:regularity} holds, so \Cref{lem:jacobian} gives $J_{ik}=P_{ik}-D_{ik}$ for $k\ne i$ and $J_{ii}=\sum_{k\ne i}D_{ik}$. Fix $\beta\in\B$ and a bidder $i$, and write $\rho_{ij}=\rho(\sigma,w_{ij}(\beta))$. By \Cref{lem:logit},
\[
J_{ik}=\sum_j\mu_j(\rho_{ij}-1)D^{(j)}_{ik}.
\] By \Cref{lem:rho}, every $\rho_{ij}$ lies in $[\sigma,\rho(\sigma,\omega)]$, so
\[
|\rho_{ij}-1|\le\max\{1-\sigma,\ \rho(\sigma,\omega)-1\}=1-\gamma .
\] The triangle inequality therefore gives
\[
\sum_{k\ne i}|J_{ik}|
\le\sum_j\mu_j|\rho_{ij}-1|\sum_{k\ne i}D^{(j)}_{ik}
\le(1-\gamma)\sum_j\mu_j\sum_{k\ne i}D^{(j)}_{ik}
=(1-\gamma)J_{ii} .
\] All $q_{kj}$ are positive, so $J_{ii}>0$. This is \eqref{eq:uniform-thickness} at every $\beta\in\B$, that is, $(D^+_\gamma)$. If $\rho(\sigma,\omega)\le1$, then every $\rho_{ij}\le1$, and the display for $J_{ik}$ gives $J_{ik}\le0$.

\Cref{thm:br,thm:dd} now apply directly. For \Cref{thm:sa}, two hypotheses remain. First, $\mu$ must be a probability measure. If $M=\sum_j\mu_j\ne1$, divide the masses and the budgets by $M$. This divides $S$, $J$ and $B$ by $M$ and leaves $\B$, the win shares and $(D^+_\gamma)$ unchanged. Second, the price at the unpaced profile must have a finite second moment. On each type, the price exceeds $x$ only if two distinct values exceed $x$, and $\Pr(e^{a_k+\sigma\varepsilon_k}>x)\le e^{a_k/\sigma}x^{-1/\sigma}$. So $\Pr(p>x)=O(x^{-2/\sigma})$, and since $2/\sigma>2$, the price has a finite second moment.
\end{proof}

\proplarge*

\begin{proof}
Write
\[
w_{ij}=\Bigl[1+\sum_{k\ne i}q_{kj}/q_{ij}\Bigr]^{-1},\qquad
q_{kj}/q_{ij}=e^{(\beta_k+a_{kj}-\beta_i-a_{ij})/\sigma}.
\] On $\B$ we have $\beta_k\ge\ell_k$ and $\beta_i\le0$, so $q_{kj}/q_{ij}\ge e^{(a_{kj}+\ell_k-a_{ij})/\sigma}$, which gives \eqref{eq:share-bound}. Under the counting hypothesis, at least $N$ of these terms are at least $e^{-W/\sigma}$, and the rest are positive. The bracket is therefore at least $1+Ne^{-W/\sigma}$, which gives the bound on $\omega$.

By \Cref{lem:rho}, $\rho(\sigma,\omega)<2$ if and only if $\omega<w_2(\sigma)$. The displayed condition on $N$ is equivalent to $(1+Ne^{-W/\sigma})^{-1}<w_2(\sigma)$, so \Cref{thm:logit} applies. As $N\to\infty$, the bound on $\omega$ tends to $0$, so $\rho(\sigma,\omega)\to\sigma$ by \Cref{lem:rho}. Once $\rho(\sigma,\omega)\le2-\sigma$, the certified contraction factor $\max\{1-\sigma,\rho(\sigma,\omega)-1\}$ equals $1-\sigma$. The bound depends only on $N$, $W$ and $\sigma$, not on $n$.
\end{proof}

\subsection{Proofs for \texorpdfstring{\Cref{sec:finite}}{Section 7}}

\lemvc*

\begin{proof}
Write $\mu_m=m^{-1}\sum_{l=1}^m\delta_{\theta_l}$ for the empirical item distribution. We reduce spend approximation to uniform approximation of the probabilities of certain sets of items.

\emph{Step 1: spend as an integral of event probabilities.} For $\beta\le0$, every price lies in $[0,\bar v]$. The identity $p=\int_0^{\bar v}\ind\{p>s\}\dd s$ gives
\[
\hat S_i^{>}(\beta)-S_i(\beta)=\int_0^{\bar v}(\mu_m-\mu)(A_{\beta,s})\dd s,
\] where, under the convention that counts only unique highest bids,
\[
A_{\beta,s}=\{i\text{ wins}\}\cap\{p>s\}
=\bigcap_{k\ne i}\{v_i>e^{\beta_k-\beta_i}v_k\}\ \cap\ \bigcup_{k\ne i}\{v_k>se^{-\beta_k}\}.
\] The intersection says that $i$ beats every rival. Given that event, the union says that the highest rival bid, which sets the price, exceeds $s$. Replacing $>$ by $\ge$ in the intersection gives the convention that also counts ties. Both conventions have fluid spend $S_i$, since positive-price ties are $\mu$-null on $U$: by \Cref{ass:smooth}, $S_i$ is continuous on $U$, whereas a positive mass of positive-price ties involving bidder $i$ would make $S_i$ jump upward as $\beta_i$ increases.

\emph{Step 2: counting event patterns.} On any $m$ sample points, each threshold family $\{v_i>cv_k\}_{c>0}$ or $\{v_k>c\}_{c>0}$ is nested, so it produces at most $m+1$ distinct subsets. There are $n-1$ families of each kind. Allowing all their thresholds to vary independently only enlarges the class $\mathcal A_i$ of events, so its shatter coefficient satisfies $s(\mathcal A_i,m)\le(m+1)^{2(n-1)}$. The same bound holds for the convention that counts ties.

\emph{Step 3: uniform concentration.} The Vapnik--Chervonenkis inequality \citep[Thm.~12.5]{devroye1996probabilistic} gives
\[
\Pr\Bigl\{\sup_{A\in\mathcal A_i}|\mu_m(A)-\mu(A)|>u\Bigr\}\le8s(\mathcal A_i,m)e^{-mu^2/32}.
\] A union bound over $n$ bidders and two conventions bounds the failure probability by $16n(m+1)^{2(n-1)}e^{-mu^2/32}$. Taking
\[
u=\sqrt{\frac{32\bigl(2n\log(m+1)+\log(16n/\varrho)\bigr)}{m}}
\] makes this probability at most $\varrho$. On the complementary event, Step 1 bounds every spend error by $\bar vu=\varepsilon_m$, uniformly over $U\cap\R^n_{\le0}$.
\end{proof}

\thmfinite*

\begin{proof}
We translate spend error into best-response error, iterate the resulting perturbed contraction, and rescale the final multipliers.

\emph{Step 1: one finite and one fluid best response.} Fix $i$ and $\beta_{-i}\in\B^+_{-i}$, suppress the rival profile, and write $b=\BR_i(\beta_{-i})\in[\ell_i,0]$.

For the lower bound, let $x_0=b-\zeta$, which lies in $[\ell_i-1,0]$ because $\zeta\le1$. We claim $S_i(x_0)\le B_i-\varepsilon$. Otherwise, monotonicity gives $S_i\ge B_i-\varepsilon\ge B_i/2$ on $[x_0,b]$, and (F3) gives
\[
S_i(b)\ge S_i(x_0)+\underline J(b-x_0)>B_i-\varepsilon+\underline J\zeta=B_i,
\] contradicting feasibility of the fluid best response. Closeness then gives $\hat S_i^{>}(x_0)\le S_i(x_0)+\varepsilon\le B_i$, so $x_0$ is feasible for the finite problem and $\widehat{\BR}_i\ge b-\zeta$.

For the upper bound, there is nothing to prove if $b=0$. Otherwise $S_i(b)=B_i$. For any $x\in(b+\zeta,0]$, monotonicity keeps spend at least $B_i$ on $[b,x]$, so (F3) gives $S_i(x)\ge B_i+\underline J(x-b)>B_i+\varepsilon$. Closeness gives $\hat S_i^{>}(x)>B_i$, ruling out every such $x$. Hence $\widehat{\BR}_i\le b+\zeta$, which proves (i).

\emph{Step 2: the perturbed contraction.} Part (i) implies $\widehat{\BR}_i\ge\ell_i-\zeta$, so all iterates started at zero remain in $\prod_i[\ell_i-\zeta,0]\subseteq\B^+$. Let $e_t=\norm{\beta^t-\beta^*}_\infty$. Since $\BR(\beta^*)=\beta^*$ and $\BR$ is a $(1-\gamma)$-contraction of $\B^+$,
\[
e_{t+1}\le\norm{\widehat{\BR}(\beta^t)-\BR(\beta^t)}_\infty+\norm{\BR(\beta^t)-\BR(\beta^*)}_\infty\le\zeta+(1-\gamma)e_t .
\] Unrolling, and using $e_0=\norm{\beta^*}_\infty\le\max_i\abs{\ell_i}$,
\[
e_t\le(1-\gamma)^te_0+\zeta\sum_{s=0}^{t-1}(1-\gamma)^s\le(1-\gamma)^t\max_i\abs{\ell_i}+\frac{\zeta}{\gamma},
\] which proves (ii).

\emph{Step 3: spend at the finite iterate.} For $t\ge t_0$, $(1-\gamma)^t\le e^{-\gamma t}$ makes the transient term in (ii) at most $\zeta/\gamma$, so $r=e_t\le2\zeta/\gamma$. Write $\varepsilon'=\varepsilon+\mathsf Lr$. The spend $\hat S^x_i(\beta^t)$ lies between $\hat S_i^{>}(\beta^t)$ and $\hat S^{\ge}_i(\beta^t)$, both within $\varepsilon$ of $S_i(\beta^t)$, and $S_i$ is $\mathsf L$-Lipschitz on the convex box $\B^+$. Hence
\[
|\hat S^x_i(\beta^t)-S_i(\beta^*)|\le|\hat S^x_i(\beta^t)-S_i(\beta^t)|+|S_i(\beta^t)-S_i(\beta^*)|\le\varepsilon+\mathsf Lr=\varepsilon'.
\] Since $S_i(\beta^*)\le B_i$, it follows that $1\le\Lambda\le1+\varepsilon'/B_{\min}$. Subtracting $\log\Lambda$ from every log multiplier preserves all bid rankings and divides every payment by $\Lambda$. Thus items still go to highest bidders and remain fully allocated, so (a$'$) holds with $\delta=0$ and (b) holds; by the definition of $\Lambda$, every rescaled spend is at most its budget, which is (c).

\emph{Step 4: no unnecessary pacing.} Set $a=\varepsilon'/B_{\min}$, so that $\phi=2a+r$. We check (d$'$) according to whether bidder $i$ is paced at the fluid equilibrium.
\begin{itemize}[leftmargin=1.5em]
\item If $\beta^*_i<0$, then $S_i(\beta^*)=B_i$. For $a\le1$, the rescaled spend per item is at least
\[
\frac{B_i-\varepsilon'}{1+a}\ge B_i\frac{1-a}{1+a}\ge B_i(1-2a)\ge B_i(1-\phi).
\] For $a>1$, the bound $B_i(1-\phi)$ is negative and holds trivially. In either case the premise of (d$'$) fails.
\item If $\beta^*_i=0$, then $\beta^t_i\ge-r$, so the rescaled multiplier satisfies
\[
\frac{e^{\beta^t_i}}{\Lambda}\ge\frac{e^{-r}}{1+a}=e^{-r}-\frac{e^{-r}a}{1+a}\ge1-r-a\ge1-\phi,
\] and (d$'$) holds whether or not bidder $i$ underspends.
\end{itemize}
Finally, substituting $r\le2\varepsilon/(\gamma\underline J)$ into $\phi=2(\varepsilon+\mathsf Lr)/B_{\min}+r$ gives
\[
\phi\le\frac{2\varepsilon}{B_{\min}}\Bigl(1+\frac{2\mathsf L}{\gamma\underline J}\Bigr)+\frac{2\varepsilon}{\gamma\underline J}=\frac{2\varepsilon\chi_\mu}{\tilde v}.
\qedhere
\]
\end{proof}

\corckk*

\begin{proof}
By (F1) and \Cref{lem:vc}, with probability at least $1-\varrho$ the instance is $\varepsilon_m$-close to $\mu$ on $\B^+\subseteq U\cap\R^n_{\le0}$. The condition on $m$ is equivalent to $2\varepsilon_m\chi_\mu/\bar v\le n^{-c}$. It also gives $\varepsilon_m\le B_{\min}/2$ and $\varepsilon_m\le\gamma\underline J/2\le\underline J$, since $\chi_\mu\ge\bar v/B_{\min}$, $\chi_\mu\ge\bar v/(\gamma\underline J)$ and $n^{-c}\le1$. \Cref{thm:finite} with $\varepsilon=\varepsilon_m$ then gives a $(0,n^{-c})$-approximate \sppe for every $t\ge t_0$. For the iteration count, $n\ge2$ and $\varrho\le1$ give $\varepsilon_m\ge\bar v\sqrt{32\log32/m}\ge\bar v/\sqrt m$, so with $\zeta=\varepsilon_m/\underline J$,
\[
t_0\le\Bigl\lceil\gamma^{-1}\Bigl(\tfrac12\log m+\log^+\bigl(\gamma\underline J\max_i\abs{\ell_i}/\bar v\bigr)\Bigr)\Bigr\rceil ,
\] where the $O(\cdot)$ in the statement hides the second, $\mu$-dependent term. For the cost of a round, bidder $i$'s thresholds are the ratios $\max_{k\ne i}b_k(\theta_l)/v_i(\theta_l)$. They can be read off from the top two bids on each item in $O(nm)$ total time, and sorting them costs $O(m\log m)$ per bidder.
\end{proof}

\corhard*

\begin{proof}
Run finite best responses from $\beta^0=0$ for $\mathsf P(n,m)$ rounds. After each round, rescale by $\Lambda$ and check conditions (a$'$), (b), (c) and (d$'$) of \Cref{def:approx} with parameters $(0,n^{-c})$, which takes polynomial time. For $I\in\mathcal I_{\mathsf P,c}$, \Cref{thm:finite}(iii) guarantees success by round $t_0\le\mathsf P(n,m)$, because $2\varepsilon\chi_\mu/\tilde v\le n^{-c}$. A $(0,n^{-c})$-approximate \sppe is in particular an $(n^{-c},n^{-c})$-approximate one.

The computation can be carried out exactly in rational arithmetic when values and budgets are rational. In multiplier form, each finite best response is either $1$ or a threshold $e^{\beta^t_k}v_k(\theta_l)/v_i(\theta_l)$. The bit length of the multipliers therefore grows by at most twice the input bit length per round, and stays polynomial over $\mathsf P(n,m)$ rounds. The rescaling factor $\Lambda$ is a ratio of sums of such numbers. Hence every round and every check take polynomial time.
\end{proof}

\end{document}